\documentclass[aps,pre,twocolumn,superscriptaddress]{revtex4}

\usepackage{epsfig,amsmath,amssymb,amsfonts,physics,graphicx,color,calc,epstopdf}

\usepackage[T1]{fontenc}

\usepackage{subcaption}

\usepackage{dcolumn}
\usepackage{bm}
\usepackage{graphicx}
\usepackage{xcolor}

\usepackage{comment}

\usepackage{physics}
\usepackage{tensor}
\usepackage{tikz}
\def\x{{\bf x}}
\def\xx{{\bf x}}
\def\y{{\bf y}}
\def\v{{\bf v}}

\def\vv{{\bf v}}
\def\w{{\bf w}}
\def\ww{{\bf w}}
\def\sign{{\rm  sign}}
\def\dde{{\rm d}}

\begin{document}

\title{The effective noise of Stochastic Gradient Descent}

\author{Francesca Mignacco}
\affiliation{Universit\'e Paris-Saclay, CNRS, CEA,
Institut de physique th\'eorique,  91191, Gif-sur-Yvette, France.}
\author{Pierfrancesco Urbani}
\affiliation{Universit\'e Paris-Saclay, CNRS, CEA,
Institut de physique th\'eorique,  91191, Gif-sur-Yvette, France.}

\begin{abstract}
Stochastic Gradient Descent (SGD) is the workhorse algorithm of deep learning technology.  At each step of the training phase, a mini batch of samples is drawn from the training dataset and the weights of the neural network are adjusted according to the performance on this specific subset of examples. The mini-batch sampling procedure 
introduces a stochastic dynamics to the gradient descent, with a non-trivial state-dependent noise. We characterize the stochasticity of SGD and a recently-introduced variant, \emph{persistent} SGD, in a prototypical neural network model.
In the under-parametrized regime, where the final training error is positive, the SGD dynamics reaches a stationary state and we define an effective temperature from the fluctuation-dissipation theorem, computed from dynamical mean-field theory. We use the effective temperature to quantify the magnitude of the SGD noise as a function of the problem parameters.
In the over-parametrized regime, where the training error vanishes, we measure the noise magnitude of SGD by computing the average distance between two replicas of the system with the same initialization and two different realizations of SGD noise. We find that the two noise measures behave similarly as a function of the problem parameters. Moreover, we observe that noisier algorithms lead to wider decision boundaries of the corresponding constraint satisfaction problem. 
\end{abstract}

\maketitle

\emph{Introduction --} Deep learning is a family of machine learning technologies that has transformed our computational approach in a range of contexts, from everyday-life practical applications to a set of new tools available to scientific research \cite{lecun2015deep,zdeborova2020understanding}. 
In a nutshell, deep neural networks are a special class of functions that automatically extract high-level features from data.  Along with the benefits brought by the introduction of \emph{deep} architectures with multiple neural layers \cite{Bengio2013RepresentationLA}, a huge leap forward in the significance of this technology resulted from the massive amount of classified data collected throughout the past decade.

Training a supervised learning model consists in minimizing a cost function that depends on the parameters of the network and on the dataset.  Neural networks usually operate in the over-parametrized regime, where the number of parameters can reach the order of $10^6-10^7$ and largely exceed the number of training samples, typically around $10^4-10^5$ and going up to $10^6-10^7$ only for the largest (open) datasets \citep{Wu2019TencentMA}. Minimizing a cost function in such a high-dimensional space is extremely complex already for general-purpose algorithms such as pure gradient descent (GD).  Indeed, calculating the gradient of the loss function brings a great computational burden, requiring the evaluation of the current state of the neural weights on the full training set. 
A more efficient alternative emerged with the introduction of the stochastic gradient descent (SGD) algorithm \cite{Bottou1999OnlineLA}. SGD, at variance with GD, approximates the gradient by evaluating it only on a \emph{mini batch} -- a small subset of the training set -- which is changed at each step of the dynamics. Quite surprisingly,  over-parametrized neural networks trained by SGD -- that can perfectly fit even random data -- do not incur in over-fitting on real data. Instead,  they can achieve excellent performances on previously unseen data \cite{Zhang2017UnderstandingDL}, a crucial property called \emph{generalization}. While it has commonly been observed that SGD outperforms GD in practical applications \cite{keskar2017largebatch}, theoretical results in support of this claim remain sparse \cite{abbe2020polytime,haochen2020shape} and current optimization strategies are largely based on heuristics.

The machine learning community is actively working to bridge this gap. A recent stream of works is aiming at characterizing the nature of the noise introduced by the mini-batch sampling procedure, by quantifying it and identifying its properties in order to understand how they correlate with the final generalization performance. 
From an analytic point of view, a large portion of this literature invokes the central-limit theorem (CLT) to model SGD as an approximated gradient-descent dynamics perturbed by simple Gaussian Langevin noise, with a variance depending on the size of the mini batch and on the time step, or \emph{learning rate}, $\dd t$ \cite{hu2017diffusion,li2017stochastic,cheng2020stochastic,jastrzkebski2017three}.  However,  this approach leads to a stochastic differential equation involving a hybrid, ill-defined continuous-time limit \cite{yaida2018fluctuation}, where the learning rate is sent to zero $\dd t \rightarrow 0^+$ in the dynamics, but at the same time it is kept finite in the noise variance. Moreover,  the validity of the CLT in this context has been questioned \cite{li2021validity}.  This perspective has also been challenged by a set of experiments \cite{simsekli2019tail} undermining the validity of the Gaussian approximation and suggesting that in fact the SGD noise may be responsible for L\'evy flights in the phase space of the weights during training. 

Tracking analytically the whole trajectory of the algorithm without resorting to approximations is more demanding. It has been done only in some particular cases: for linear networks trained by GD \cite{BO97,SMG13, bodin2021model}; for single-layer or two-layer networks trained by \emph{online} SGD  \cite{SS95Short,SS95,coolen2000,Sa09,GASKZ19,rotskoff2018neural,MMN18,chizat2018global}.   However, linear networks lack the expressive power of non-linear ones. In the case of online learning, there is no distinction between training and test loss, therefore the generalization gap -- the gap between train and test error -- is not defined.

Recent works \cite{Mignacco_2021,francesca2020dynamical} showed that the dynamical trajectory of SGD can be tracked analytically via dynamical mean-field theory (DMFT) from statistical physics \cite{MPV87,GKKR96, agoritsas2018out,PUZ20}, even for multi-pass SGD (i.e., when the mini batches are reused multiple times). This description is not restricted to the continuous-time limit: indeed, the discretized version of the DMFT equations is able to capture the dynamics of SGD also at finite learning rate, provided the weight increments between two updates do not involve higher orders terms than $\mathcal{O}(\dd t)$.  

In this work, we apply DMFT to characterize the late-time dynamics of SGD and quantify its algorithmic noise.  In order to decouple the effect of SGD noise on optimization from that of network architecture and data structure, we focus on a single-layer network and a simple loss landscape. In particular, we consider a prototypical supervised-classification problem and we integrate the DMFT equations in this setting up to times when the dynamics has either reached a stationary state or stopped.  We highlight the difference between these two possible scenarios. Indeed, following the analogy with constraint-satisfaction problems \citep{FPSUZ17}, the parameters space can be split into two regions: the under-parametrized or \emph{unsatisfiable} (UNSAT) phase, where the network cannot achieve zero training error and the dynamics goes to a stationary state, and the over-parametrized or \emph{satisfiable} (SAT) phase, where the dynamics stops at one solution with zero training error.
In the UNSAT phase, computing the correlation and response functions of the network weights, we characterize the stationary state by defining an effective temperature $T_{\rm eff}$ from the fluctuation-dissipation relation. From this relation, we then extrapolate numerically the value of $T_{\rm eff}$, which relates correlation and response at stationarity and quantifies the magnitude of SGD noise as a function of the problem hyper-parameters. 

In the SAT phase, the extrapolated temperature approaches zero at large times. This result aligns with the intuitive picture that SGD implements a self-annealing procedure while navigating the loss landscape \cite{Fenge2015617118}.  In order to assess the noise magnitude in the SAT phase, we introduce an alternative measure that we can access both analytically -- via DMFT -- and from numerical simulations.  We consider the dynamics of two copies of the system, starting from the same initial condition but subjected to two different realizations of the stochastic noise, i.e., two different histories of mini-batch sampling. We then track the average distance $d(t)$ between these two trajectories at time $t$ as they evolve in the weight space and when they finally land on a border of the zero-training-error region.  We use this distance to quantify the noise of the SGD algorithm as a function of the problem hyper-parameters. The consistency of the two definitions in the UNSAT phase is shown in Fig. \ref{fig:distance_unsat} of Appendix \ref{appendix:distance_unsat}. Remarkably, we show that a higher noise is associated to a smaller fraction of support vectors at the end of the training and therefore to a more robust solution \cite{xu2009robustness}. 

We investigate the role of the various hyper-parameters in the SAT and UNSAT phase, which could provide theoretically-informed guidance for practical implementation.  We find a qualitative agreement in the behavior of the two different measures of noise magnitude, $T_{\rm eff}$ and $d(t)$, as a function of the hyper-parameters.

\section{Model and Methods} 
We study a prototypical binary classification problem in high dimensions.  The network is presented with a set of $M$ training examples in dimension $N$, $X=(\x_1, .., \x_M)^\top\in \mathbb{R}^{M\times N}$, and the corresponding binary labels $\y =(y_1,...,y_M)^\top\in \{-1,+1\}^M$. The goal is to learn the classification rule.  We specify a generative model for the data, that we assume drawn from a binary mixture of Gaussian vectors centered in the unknown vectors $\pm\v^*\in\mathbb{R}^N$.  The training samples ${\bf x}_\mu$ are independently drawn from $\x_\mu\sim\mathcal{N}(y_\mu\v^*/\sqrt{N},\Delta\,I_N)$, where $I_N$ stands for the $N-$dimensional identity matrix. The data noise is tuned by the parameter $\Delta>0$.  The labels are set to $y_\mu = \pm 1$ with equal probability.  Due to the statistical isotropy of the samples, without loss of generality we can take $\v^*=(1,1,..,1)\in \mathbb{R}^N$. We consider the thermodynamic limit, where $M,N\rightarrow\infty$ at fixed sample complexity $\alpha=M/N\sim\mathcal{O}(1)$. In this limit, if the ratio $\alpha$ is smaller than a critical threshold $\alpha_c(\Delta)$,  then the two Gaussian clouds can be separated by an hyperplane, as studied in detail in \cite{mignacco2020role}.  Therefore,  a single-layer neural network that estimates the labels according to the linear rule $\hat y _\mu(\w)=\sign(\w^\top\x_\mu/\sqrt{N})$ is well-suited to perform the classification. The region of sample complexity $\alpha<\alpha_c$ corresponds to the SAT phase,  while $\alpha>\alpha_c$ to the UNSAT phase. The optimal performance is achieved by setting ${\bf w}={\bf v^*}$.

The weight vector $\w \in \mathbb{R}^N$ is learned via the empirical risk minimization \cite{NIPS1991_ff4d5fbb} of the loss
\begin{equation}
\mathcal{L}(\w)=\sum_{\mu=1}^M\ell\left(\frac{y_\mu}{\sqrt{N}}\x_\mu^\top \w\right)+\frac \lambda 2\Vert \w\Vert_2^2,
\label{eq:loss}\end{equation}
where $\ell(\cdot)$ is a cost function accounting for the per-sample error. In what follows, we will always consider the \emph{squared hinge} cost function $\ell(h)=(h-\kappa)^2\,\Theta(\kappa-h)/2$, with $\kappa>0$ and $\Theta(\cdot)$ indicating the Heaviside step function.  We note that this particular loss is zero as soon as all the samples are correctly classified with a robustness ensured by the \emph{margin} $\kappa$, i.e., $y_\mu\x_\mu^\top \w/{\sqrt{N}}>\kappa$ for all the samples $\mu\in\{1,\ldots M\}$. For this reason, in the SAT phase, the dynamics stops as soon as a solution is found. This choice is not restrictive, indeed in real implementations the dynamics is usually stopped as soon as the training loss goes below some threshold value. As customary in practical applications, we have added a soft constraint on the $\ell_2-$norm of the weights, called \emph{ridge regularization} \cite{ridge_reg}. For simplicity, the strength $\lambda\geq 0$ of this regularization will be fixed during training. Note that the SAT phase is realized only at $\lambda=0$. \eqref{eq:loss} defines an optimization problem over a landscape characterized by the quenched disorder coming from the training set.
\begin{figure}[t!]
\centering
\includegraphics[scale=0.23]{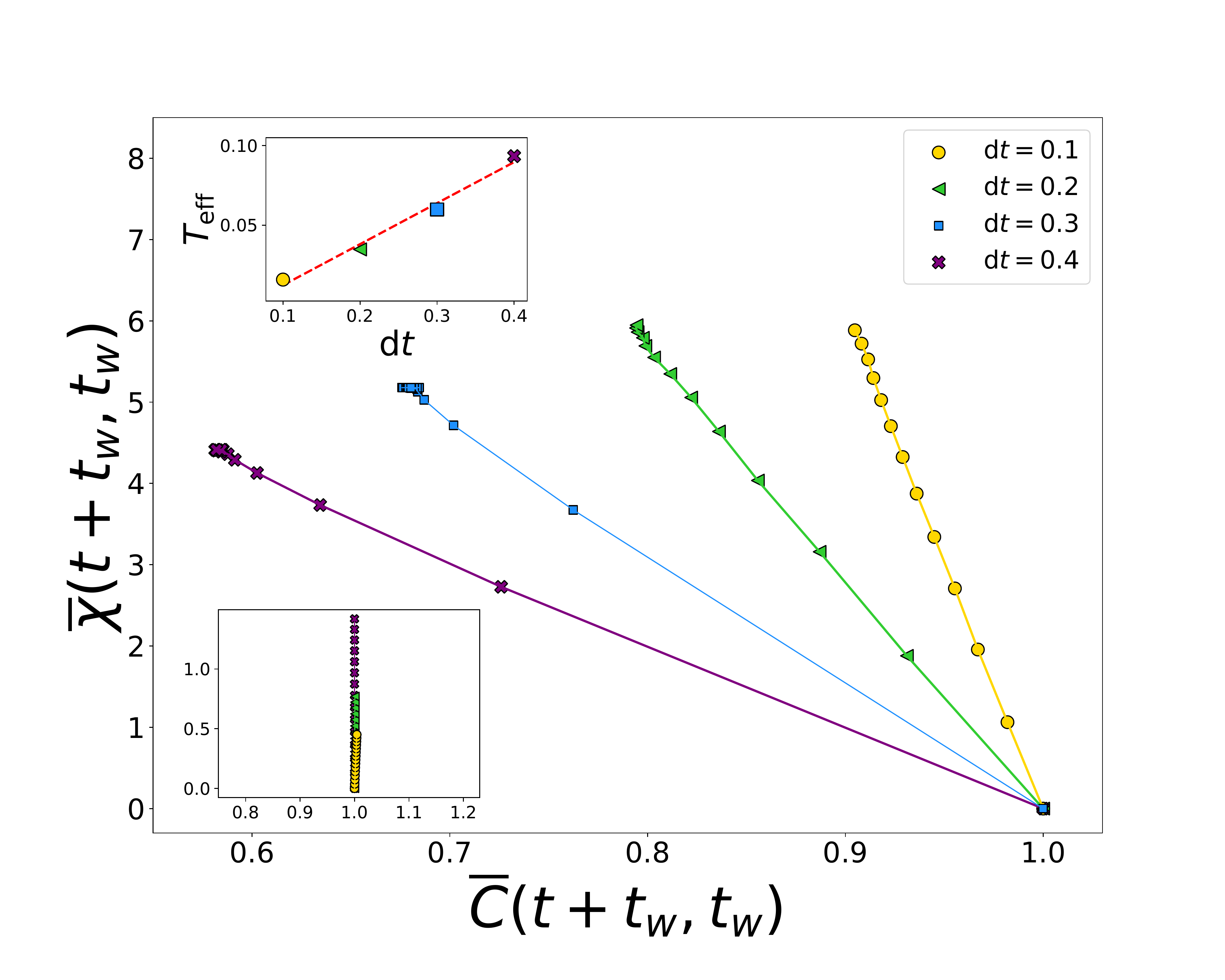}\vspace{-.4em}
\centering
\includegraphics[scale=0.23]{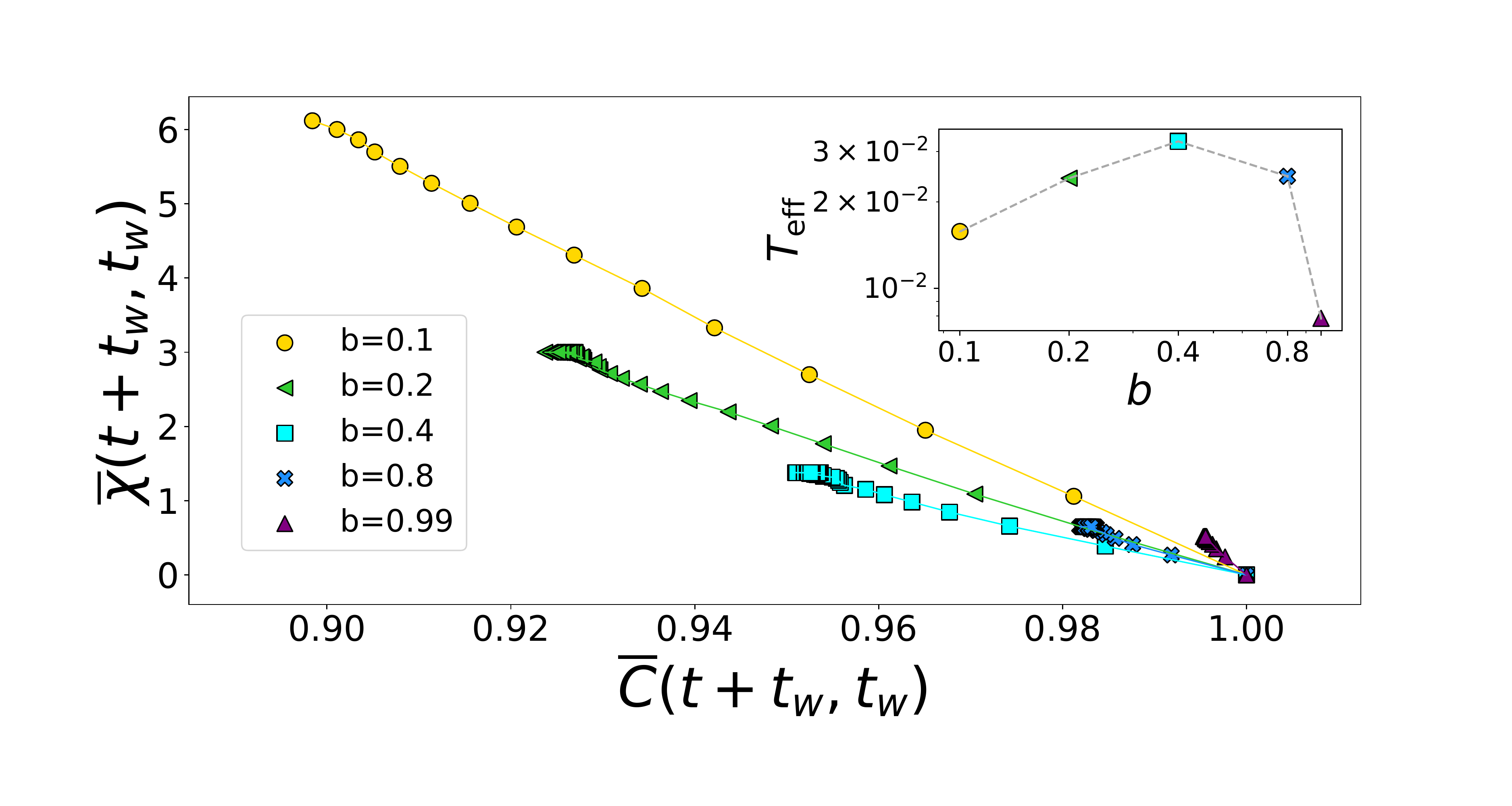}
\caption{FDT plot for vanilla-SGD. \emph{Top panel}: the different curves represent different choices of learning rate $\dde t =0.1,0.2,0.3,0.4$. The main plot is obtained in the UNSAT phase ($b=0.1,\alpha=6, \Delta=1, \lambda=1$), while the lower inset depicts the SAT phase ($b=0.5,\alpha=2$, $\Delta=0.5$, $\lambda=0$). In the upper inset, we plot the behavior of the effective temperature as extracted from the main plot.
\emph{Bottom panel}: the same analysis at fixed learning rate $\dd t=0.1$ and different batch sizes $b=0.1,0.2,0.4,0.8,0.99$.
\label{fig1} }
\end{figure}
We will study optimization via the SGD algorithm, whose weight updates can be generically written as
\begin{equation}
\begin{split}
\w(t+\dd t)=\w(t)\\-\dd t\left[\sum_{\mu=1}^Ms_\mu(t)\,\nabla_{\w}\ell\left(\frac{y_\mu}{\sqrt{N}}\x_\mu^\top \w(t)\right)+\lambda \w(t)\right]\\
:=\w(t)-\dd t \, \tilde\nabla^{{\bf s}(t)}_{\w}\mathcal{L}(\w (t)) ,
\end{split}\label{eq:weight_update}
\end{equation}
where the binary vector ${\bf s}(t)\in\{0,1\}^M$ accounts for the sampling procedure at each time step, and we have defined the approximated gradient via the operator $\tilde\nabla^{{\bf s}(t)}_{\w}$, that we will also indicate component-wise via the notation $\{\tilde\partial_{w_j}^{{\bf s}(t)}\}_{j=1}^N$.  According to the different choices of the process $s_\mu(t)$, \eqref{eq:weight_update} describes different specific cases. First,  by setting $s_\mu(t)\equiv 1$ for all samples $\mu\in\{1,\ldots M\}$ and all times $t\geq 0$, we recover the GD algorithm.  In the high-dimensional limit, the vanilla-SGD algorithm -- in the case of extensive mini batches, sampled with replacement -- corresponds to drawing $s_\mu(t)$ independent identically distributed (i.i.d.) from a Bernoulli distribution of probability ${\rm Prob}\left(s_\mu(t)=1\right)=b\in(0,1]$.  In the following, we will refer to $b$, i.e., the fraction of samples in the trainining mini-batch, as \emph{batch size}. We also consider the \emph{persistent}--SGD (p--SGD) algorithm, introduced in \cite{francesca2020dynamical} in order to establish a well-defined continuous-time limit of the SGD dynamics. In this case, $s_\mu(t)$ follows a two-state Markov jump process with exponentially-distributed transition times, and inhomogeneous switching rates: $r^{s_\mu(t)}_{0\rightarrow 1}=1/\tau$ (``activation'' rate), $r^{s_\mu(t)}_{1\rightarrow 0}=(1-b)/b\tau$ (``deactivation'' rate). The value of $\tau>0$ indicates the average time spent by each sample out of the training mini-batch, and we will call it \emph{persistence time}. The average time spent in the training mini batch by each sample is $\tau b/(1-b)$. In the limit $\tau\approx \dde t/b$ we recover the vanilla-SGD algorithm.
It is interesting to notice the analogy with both driven and active matter systems, where the passive dynamics of degrees of freedom is also dressed with local and independent forcing variables (strain deformation in driven systems \cite{nicolas2018deformation} and self-propulsion in active matter \cite{marchetti2013hydrodynamics}). The main difference between active matter and SGD is that in the former each variable has its own forcing field, while in the latter the evolution of each degree of freedom is influenced by all forcing variables. 
\begin{figure}[t!]
\centering
\includegraphics[scale=0.25]{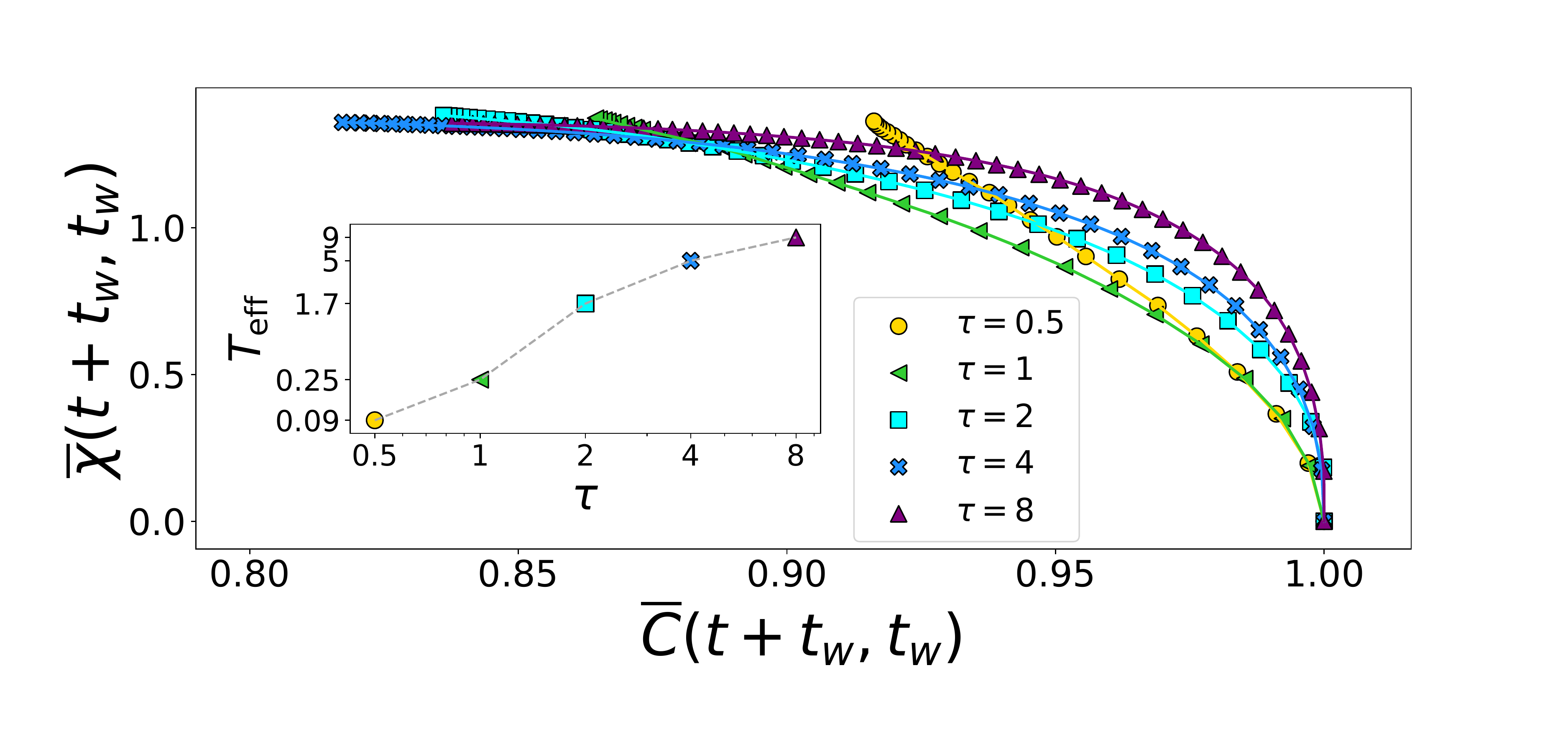}\vspace{-.7em}
\centering
\includegraphics[scale=0.25]{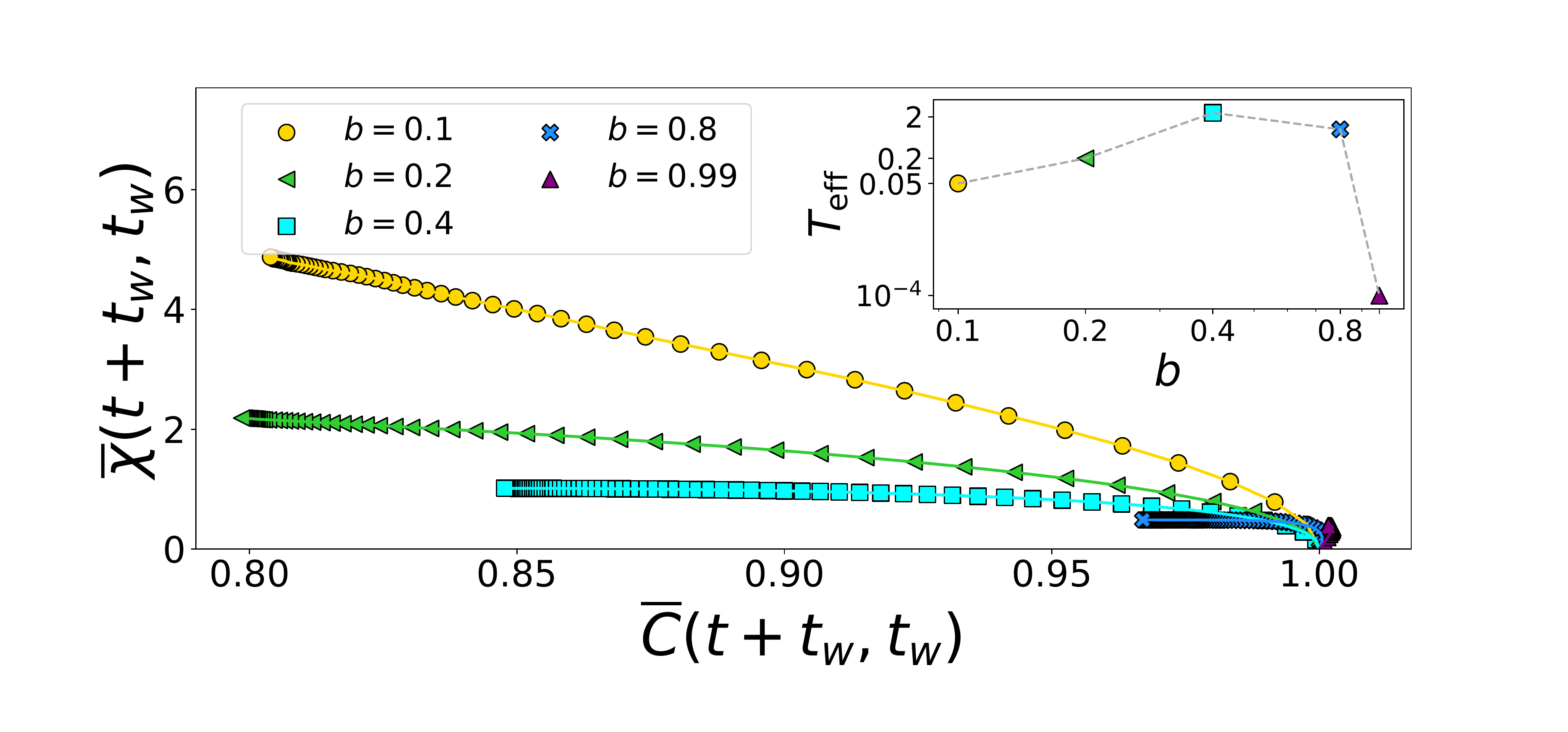}
\caption{\label{fig2}FDT plot for p-SGD. We consider $\alpha=8$, $\Delta=1$, $\lambda=1$, where the classification problem is UNSAT and we use $\dd t=0.05$.  We consider different persistence times $\tau=0.5,1,2,4,8$ and fixed $b=0.3$ (top panel), and different batch sizes $b=0.1,0.2,0.4,0.8,0.99$ and fixed $\tau=2$ (bottom panel). The insets display the effective temperature, numerically estimated, as a function of the persistence time (top panel) and batch size (bottom panel).}
\end{figure}

We use the DMFT equations derived in \cite{francesca2020dynamical} to track the dynamics of the three algorithms listed above. Since the weight increments $w_j(t+\dd t)-w_j(\dd t)$ are of order $\mathcal{O}(\dd t)$, even vanilla SGD can be exactly tracked at finite learning rate. We integrate the DMFT equations via a numerical iterative method as described in \cite{francesca2020dynamical}.

The main quantities of interest for our analysis are the dynamical correlation function
\begin{equation}
C(t,t')=\frac{1}{N}{\bf w}(t)^\top {\bf w}(t'),
\end{equation}
and the linear response function
\begin{equation}
R(t,t')=\lim_{\{ H_j\to 0\}}\frac{1}{N}\sum_{j=1}^N \frac{\delta w_j(t)}{\delta H_j(t')},
\label{def_R}
\end{equation}
where $H_j$ is an infinitesimal local field coupled to the $j^{\rm th}-$weight. In other words, one considers the dynamics where $\mathcal{L}(\w (t)) $ is shifted as $\mathcal{L}(\w (t)) -\sum_i H_i(t) w_i$. In the high-dimensional limit, these two-point functions concentrate to a deterministic value. 

In generic equilibrium stochastic processes, correlation and response are related by the fluctuation-dissipation theorem (FDT) \cite{cugliandolo2011effective}:
\begin{equation}
R(t,t')=-\frac{1}{T}\partial_{t}C(t,t')\,\Theta(t-t'),\label{fdt}
\end{equation}
where $T$ indicates the equilibrium temperature that enters in the stationary Gibbs measure. However, for a generic stochastic process that may be out of equilibrium, FDT does not hold and the stationary state is not given by the Gibbs measure. However one can define an effective temperature via \eqref{fdt}, which in general will be a function of $t$ and $t'$. This concept has proven useful across a variety of systems, ranging from glasses \cite{cugliandolo2011effective} to, more recently, active matter \cite{loi2008effective,berthier2013non} \footnote{We note that the actual meaning of $T_{\rm eff}$ extracted from the FDT theorem as a thermodynamic temperature is not granted and this is an open question in generic out-of-equilibrium systems, see \cite{cugliandolo2011effective} and \cite{loi2008effective} for more details. We do not address this issue here and use the definition of the effective temperature from FDT as a way to measure the magnitude of the noise of SGD in the stationary state. }.
It is convenient to work with the integrated response:
\begin{equation}
\chi(t,t')=\int_{t'}^t \dd s \, R(s,t'),
\end{equation}
that can be computed from the DMFT equations. By integrating both sides of \eqref{fdt}, we obtain, for $t\geq t'$,
\begin{equation}
\begin{split}
\bar{\chi}(t,t')&=\frac{1}{T}\left(1-\bar{C}(t,t')\right),\label{eq:effFDT}
\end{split}
\end{equation}
where we have defined $\bar \chi (t,t')=\chi(t,t')/C(t',t')$ and $\bar C (t,t')=C(t,t')/C(t',t')$.  At equilibrium, \eqref{eq:effFDT} implies that the parametric plot of $\bar\chi$ versus (vs) $\bar C$ gives direct access to the equilibrium temperature. 

We apply the same construction to the SGD learning dynamics.
We \emph{define} the effective temperature $T_{\rm eff}$ by plotting parametrically the rescaled integrated response $\bar \chi$ vs the rescaled correlation $\bar C$ and measuring the slope of this function in the stationary state, meaning for large $t'$ and $t$ and large time difference $t- t'$. We dub the corresponding plot as the FDT plot \cite{cugliandolo2011effective}. The details of this procedure are discussed in Appendix \ref{appendix:teff}. We find that this slope is essentially constant for vanilla-SGD, meaning that in the stationary state the algorithm is characterized by an effective FDT with a well-defined effective temperature that we can compute. We implement a similar procedure for p-SGD, unveiling an interesting difference with respect to vanilla-SGD. Indeed, p-SGD has a constant effective temperature in the stationary state only when the difference $t-t'$ is larger than the typical minibatch decorrelation time  while for shorter timescales one gets a smooth continuous function which ends with an infinite slope at $\overline C=1$. This behavior is discussed in greater detail in the next section and in Appendix \ref{appendix:teff}.
\begin{figure}[t!]
\includegraphics[scale=0.23]{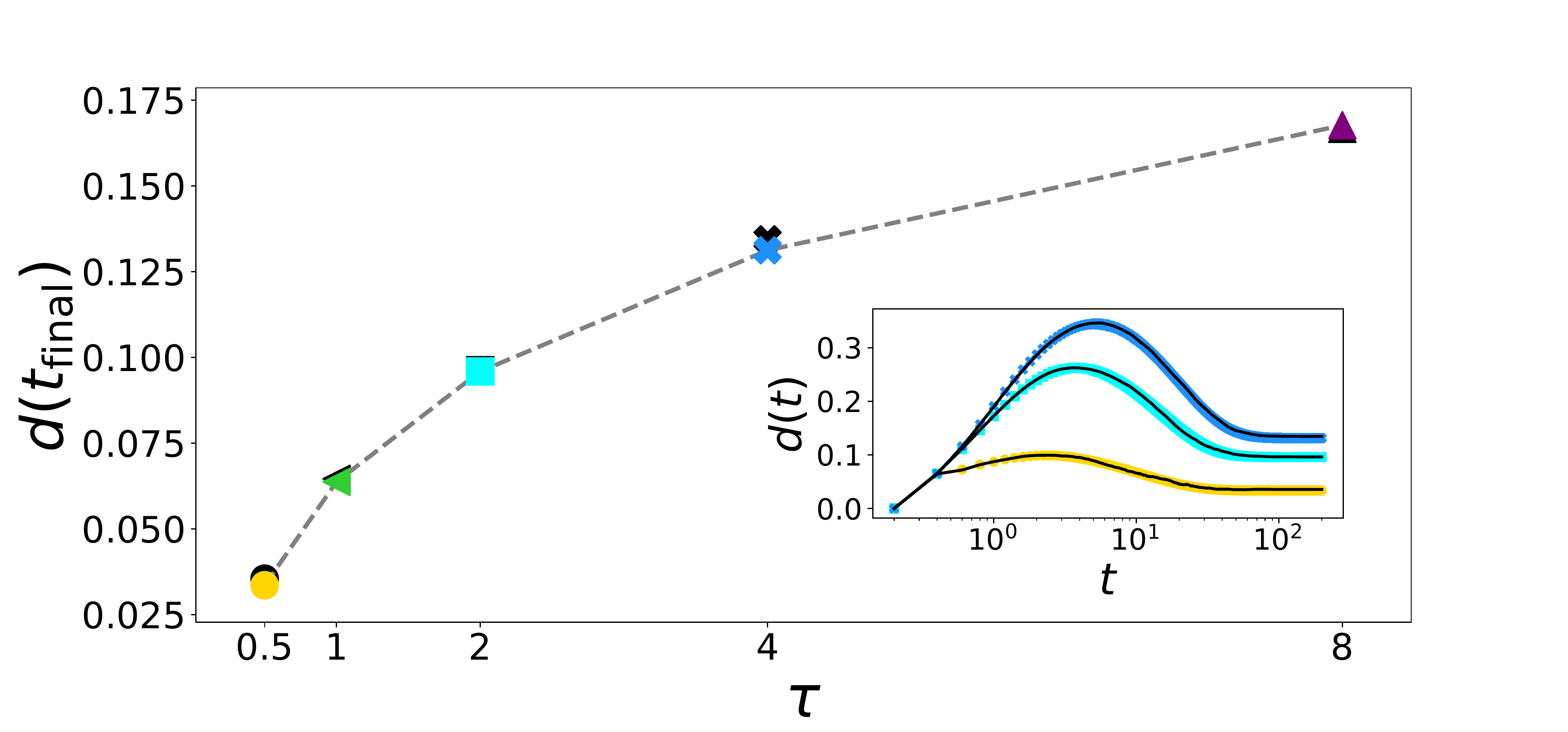}\vspace{-1mm}
\includegraphics[scale=0.23]{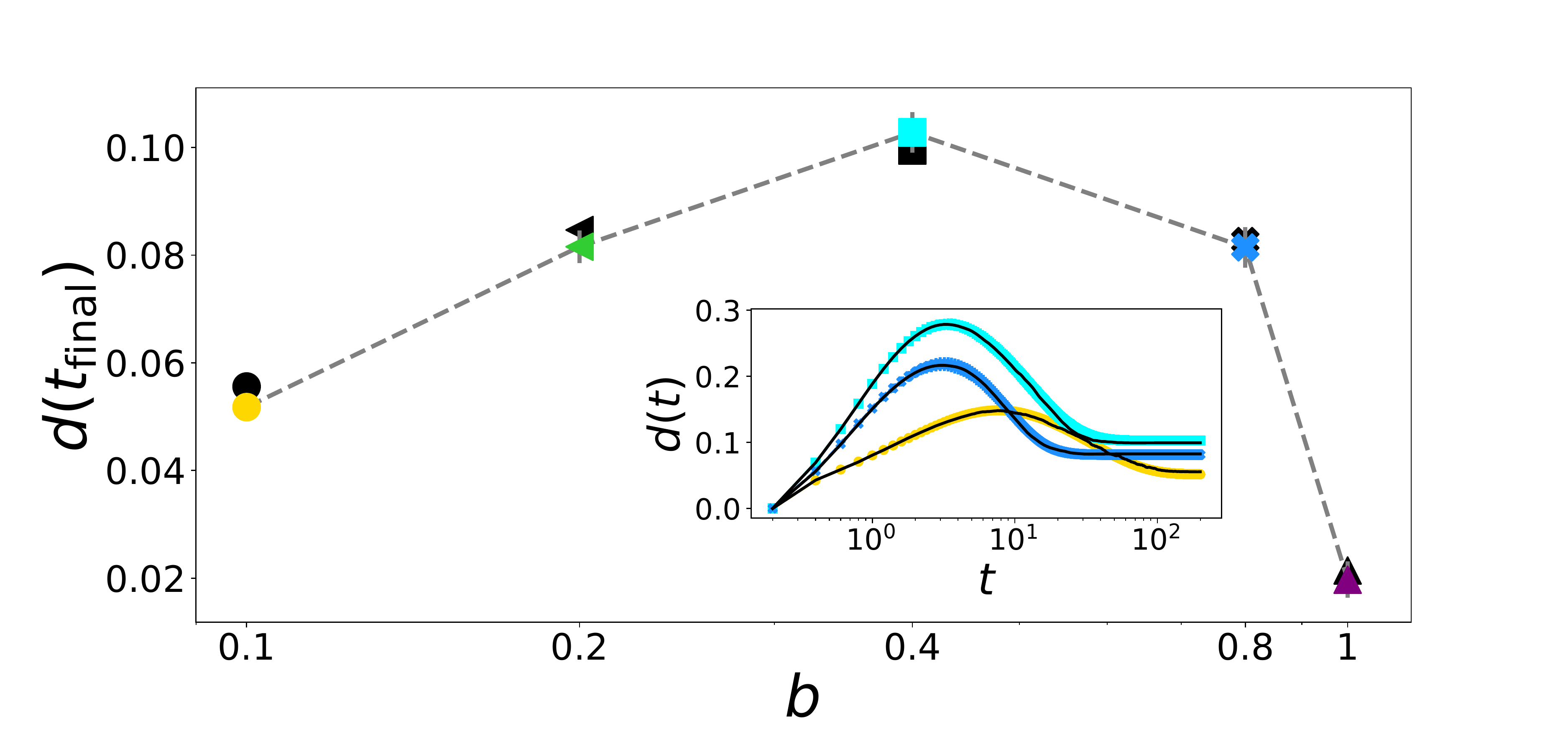}\vspace{-1mm}
\caption{Average final distance $d(t_{\rm final})$ between two copies of the p-SGD dynamics. \emph{Top panel}: We plot $d(t_{\rm final})$ as a function of the persistence time $\tau$ at fixed batch size $b=0.3$. In the inset we show the time evolution of the distance $d(t)$ from numerical simulations at $\tau=0.5$ (yellow), $\tau=2$ (cyan), $\tau=4$ (blue). \emph{Bottom panel}: We plot $d(t_{\rm final})$ as a function of the batch size at fixed persistence time $\tau=2$. In the inset we show the time evolution of the distance $d(t)$ from numerical simulations at $b=0.1$ (yellow), $b=0.4$ (cyan), $b=0.4$ (blue). In both panels, the black symbols in the main plots and the black lines in the insets mark the theoretical prediction from DMFT. The learning rate is $\eta=0.2$ in both DMFT and simulations. The other parameters are fixed such that the classification problem lies in the SAT phase: $\alpha=0.5, \Delta=0.5,\lambda=0, \kappa=1$. \label{fig:SGD}}
\end{figure}

In order to quantify the noise magnitude in the SAT phase, we consider the rescaled distance (root mean square displacement) $d(t)=\Vert \w^1(t)-\w^2(t) \Vert_2/\sqrt{N}$ between two realizations $\w^1$ and $\w^2$ of the SGD and p-SGD dynamics, starting at the same initialization point $\w^1(t=0)=\w^2(t=0)$ and subjected to two different noise realizations ${\bf s}^1(t)\neq {\bf s}^2(t)$.
At the end of training, the distance $d(t=t_{\rm final})$ quantifies the spread of the solutions found by different runs of the algorithm.  This quantity can be computed analytically via DMFT (see in particular \cite{sompolinsky1988chaos,crisanti2018path, krishnamurthy2020theory} where this procedure gives access to the Lyapunov exponent of the underlying chaotic dynamics in recurrent neural networks). The starting point of the analysis is the dynamical partition function:
\begin{equation}
\begin{split}
Z_{\rm dyn}=\mathbb{E}_{\w^{(0)}}\int_{\w^1(0)=\w^2(0)=\w^{(0)}} \mathcal{D}\w^1(t)\mathcal{D}\w^2(t)\\\times\prod_{j=1}^N\prod_{ a=1,2}\delta\left(\dot w_j^a (t)+\tilde\partial^{{\bf s}^a(t)}_{w^a_j}\mathcal{L}(\w (t))\right),
\end{split}
 \label{Zdyn_replicas}
\end{equation}
that allows to compute the correlation and response functions of the coupled system of replicas, at initialization $\w^{(0)}\sim \mathcal{N}({\bf 0}, I_N\,R)$, $R>0$. We relegate the details of this computation to Appendix \ref{appendix:dmft}.
\section{Results} We first discuss our results for the UNSAT phase, where the landscape has a unique minimum in which the SGD noise induces a non-equilibrium steady state. We compute the integrated response $\chi(t+t_w,t_w)$ and the correlation function $C(t+t_w,t_w)$. We let the system evolve until a \emph{waiting} time $t_w$ such that the stationary state has been reached. Then, at fixed $t_w$, we display the FDT plot $\bar \chi (t+t_w,t_w)$ vs $\bar C(t+t_w,t_w)$, parametrized by the time shift $t$. 
Fig. \ref{fig1} summarizes our findings regarding the effective FDT for the vanilla-SGD algorithm.
For large enough $t_w$, the relation between integrated response and correlation becomes linear and we can extrapolate numerically the effective temperature via \eqref{eq:effFDT}. The FDT plot depends both on the batch size and the learning rate. The top panel of Fig. \ref{fig1} shows that for vanishing learning rate the effective temperature of SGD approaches zero.  However, for this particular problem we observe that the vanishing-learning-rate limit of SGD does not approach GD flow, as further illustrated by numerical simulations in Appendix \ref{appendix:dt_zero}.  This observation suggests that the behavior of SGD in approximately-continous time is nontrivial and worth further investigation.  Increasing the learning rate results in a noisier dynamics and a higher effective temperature. The behavior of the effective temperature with batch size is more intriguing. Indeed, when we fix the learning rate and vary the batch size we observe a non-monotonic curve.  For a batch size close to one, the dynamics tends to GD flow and the noise shrinks to zero. If the batch size is small -- which here corresponds to the limit of sub-extensive mini batches -- we again observe a decrease in the algorithmic noise. Quite surprisingly, the highest noise is attained at intermediate extensive batch sizes. 

In the SAT phase -- displayed in the bottom inset of Fig. \ref{fig1} -- the dynamics implements an automatic self-annealing procedure, and we obtain a zero effective temperature in the zero-loss region, reached at the end of training. This observation is obvious in problems where a ``lake'', i.e. a large connected set, of solutions is found at late times so that the dynamics stops. However, the way in which the self annealing is produced drastically affects the learning trajectory in more complex problems. Indeed, in problems like phase retrieval \cite{fienup1982phase} where there is no such lake of solutions, but just one global minimum (modulo some symmetry) and a proliferation of local minima, the self-annealing property appears to be crucial to achieve good generalization \cite{Mignacco_2021}. 
\begin{figure}[t!]
\centering
\includegraphics[scale=0.23]{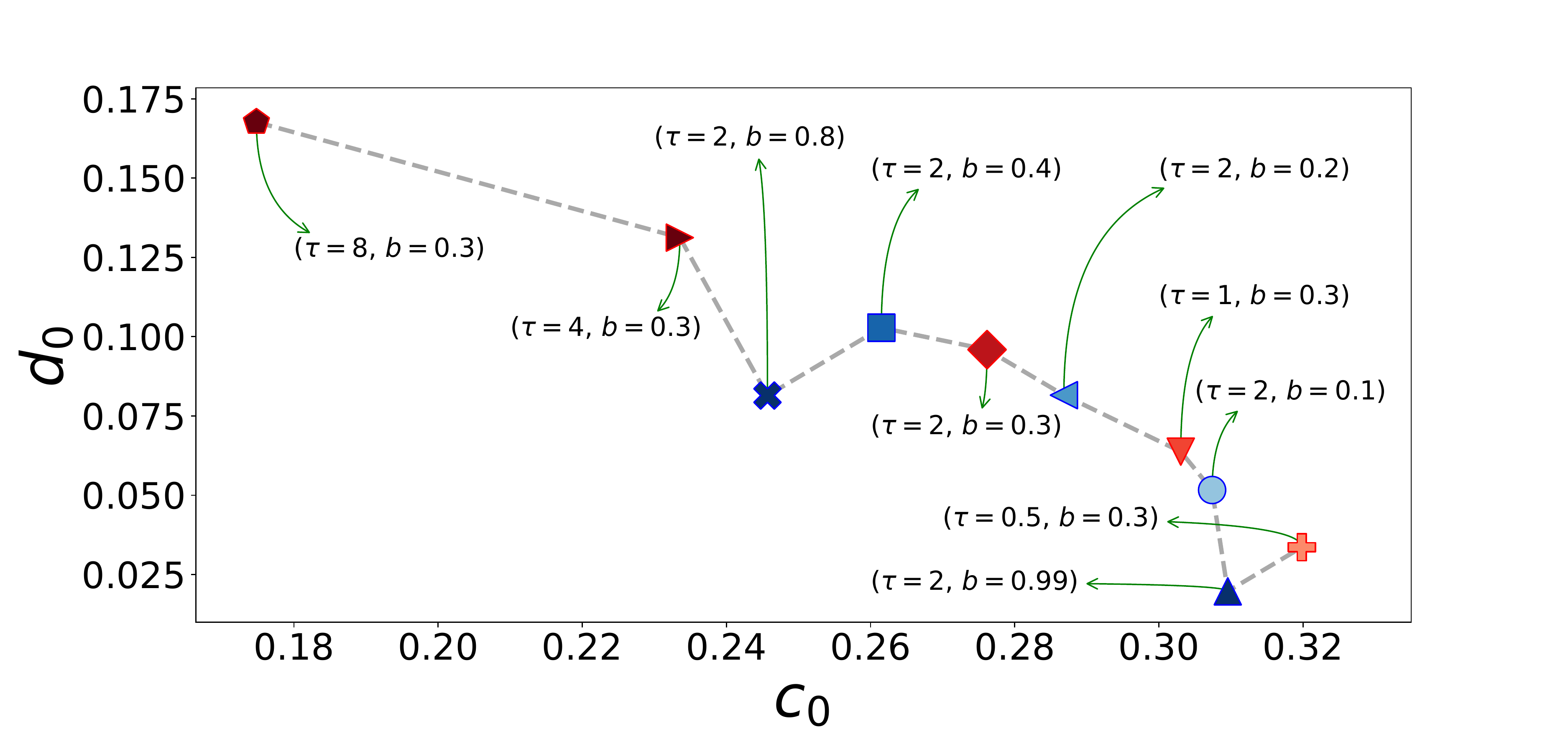}
\caption{Average distance $d_0$ between two replicas as a function of the average fraction of support vectors $c_0$, both computed as explained in Appendix \ref{appendix:sv}. Each symbol represents a different choice of the persistence time $\tau$ and the batch size $b$ in the p-SGD algorithm.  Darker colors correspond to higher values of $b$ and $\tau$.  The learning rate is fixed to $\eta=0.2$. The other parameters are fixed such that the problem lies in the SAT phase: $\alpha=0.5, \Delta=0.5,\lambda=0, \kappa=1$. \label{DC}}
\end{figure}

We turn our focus to p-SGD. In Fig. \ref{fig2}, we study the FDT plot for p-SGD changing the persistence time and the batch size. We observe that $T_{\rm eff}$ is monotonically increasing with the persistence time. The physical interpretation of this property is rather clear: if the persistence time is far from the vanilla-SGD limit, i.e., $\tau\gg \dde t/b$, the system ends up on a local minimum of a \emph{partial} loss, namely the loss function evaluated only on the samples belonging to the current mini batch. Therefore, the system is fitting well such subset of samples. 
Conversely when $t-t'\gg \tau b/(1-b)$ the dynamics has seen many mini-batches beyond the one which was there at $t'$. When a mini-batch  is renewed the system finds itself in a very atypical, random-like configuration with a large stochastic gradient. This effect produces a high effective temperature that stays constant as the FDT plot becomes linear at large $t-t'$. Note that, in the vanilla-SGD limit recovered at $\tau\approx \dde t/b$, the system never has the possibility to equilibrate in the partial loss but this comes with the effect that the stochastic gradient does not have big jumps as with p-SGD with finite $\tau$. This idea is similar to what has been done in active systems \cite{mandal2021study}, which strengthens the connection between SGD and other types of out-of-equilibrium systems.
At fixed persistence time, we observe a non-monotonic effective temperature as a function of the batch size $b$, consistently with our results for vanilla-SGD. Finally, we observe no dependence of the effective temperature on the initialization variance $R$ (see Appendix \ref{appendix:teff}), as expected given that the solution of the problem is unique in the UNSAT phase and the system reaches a stationary state which forgets the initial condition.

We now consider the characterization of the effective noise in the SAT phase, which is more interesting for practical applications since artificial feed-forward neural networks typically lie in this regime. In Fig. \ref{fig:SGD} we plot the final distance $d(t=t_{\rm final})$ between two replicas of p-SGD as a function of the batch size and learning rate. We use $d(t_{\rm final})$ to probe the algorithmic noise. Indeed, we expect that noisier regimes are associated to a higher degree of landscape exploration, resulting in a greater divergence between two replicated trajectories. We find that the behavior of $d(t_{\rm final})$ in the SAT phase mirrors the one of the effective temperature $T_{\rm eff}$ in the UNSAT phase: the effective noise increases with the persistence time and is non-monotonic with the batch size.  This is confirmed also by Fig. \ref{fig:distance_unsat} in Appendix \ref{appendix:distance_unsat}, where the same quantity is computed in the UNSAT phase.

To improve the characterization of the endpoints of the dynamics in the SAT phase, we also compute the size of the \emph{support vectors set} \cite{vapnik_sv} in the following way. In the case of GD-flow dynamics, defined in continuous time with $\dde t \rightarrow 0^+$, the endpoint of training lies on the border of the lake of solutions. We consider the number of unsatisfied constraints, i.e., misclassified samples, as it evolves with time. The long time limit of this quantity is finite, since all solutions $\w$ lying on this border marginally classify some samples, i.e., $y_\mu\,\w^\top \x_\mu=\kappa$ for some $\mu\in\{1,\ldots ,M\}$. These marginally classified samples are called \emph{support vectors}. We indicate the (rescaled) size of the set of support vectors as $c=|\{\x_\mu,\,\mu=1,\ldots,M,\;{\rm such\,that}\; y_\mu\w^\top \x_\mu=\kappa\}|/M$. In the case of p-SGD, the flow limit remains well defined, since the algorithm admits a continuous-time description \cite{Mignacco_2021}. We can therefore apply the same definition as for GD. However, while for the flow limit there are no ambiguities, in practice, integrating the dynamics requires a finite learning rate. As done in \cite{hwang2020force}, for full-batch gradient descent, we fix a threshold on the stochastic gradient, rescaled by the batch size: $\parallel \tilde \nabla _{\bf w}^{{\bf s}(t)}\mathcal{L}\parallel_2^2/bN \leq 10^{-10}$. We compute the values of $c(t)$ and $d(t)$ as soon as this threshold is reached. and we take them as the proxy for their limiting values $c_0=\lim_{t\rightarrow \infty}c(t)$, $d_0=\lim_{t\rightarrow \infty}d(t)$. This procedure is explained in Appendix \ref{appendix:sv}. 

The physical meaning of the size of the support vectors is related to the description of the local density of solutions at the endpoint of the dynamics. If $c(\w)$ is close to one, the solution $\w$ lies in a narrow corner of the solutions space. Conversely, a low value of $c$ is indicative of a wide region of the solutions space. In other words, the dynamics has landed on a ``shore'' with a wide lake of solutions just in front of it. In the latter case, one may expect the solution to be more robust to perturbations.

In Fig. \ref{DC}, we illustrate the behavior of $d_0$ as a function of $c_0$. We observe that a larger algorithmic noise leads to a smaller $c_0$. Therefore, SGD brings the system to wider regions of the lake of solutions. Note that this notion of wideness differs from the one proposed in the recent literature on wide minima, see for instance \cite{pittorino2020entropic}. Indeed, in the present case the lake of solutions is unique and the width is encoded by the number of support vectors. The smaller is $c$, the larger the density of solutions close to the endpoint of the dynamics.
\section{Discussion}
We have analyzed the nature of the stochastic noise in SGD-type algorithms in the setting of binary classification of Gaussian mixtures. We have shown that this noise can be described by an effective temperature defined through the fluctuation-dissipation theorem. In the under-parametrized regime, where the loss landscape displays a unique minimum, both vanilla-SGD and p-SGD converge to a steady state which is driven by the algorithmic noise. We have shown that the stationary state of vanilla-SGD is characterized by an effective temperature that tends to zero for vanishing learning rate, confirming that the SGD dynamics approaches GD flow for $\dde t\rightarrow 0^+$. For p-SGD, we have shown that the effective temperature increases with the persistence time, while it is non-monotonic with the batch size. In the over-parametrized regime, we have presented an alternative characterization of the magnitude of algorithmic noise. In particular, we have found that the noisier the algorithm, the smaller the fraction of support vectors at the end of the dynamics. 

These results have been derived for a simple yet paradigmatic supervised learning task. In the UNSAT phase, this setting provides the advantage that the noise captured by our analysis comes entirely from the algorithm itself since there is no other source of randomness at fixed dataset and initialization. In the following paragraph, we comment briefly on the insights that our analysis provides on the case of more complex architectures and data structures.

In more complicated settings, it is reasonable to expect that the under-parametrized regime is glassy with many local minima. This is well known in non-convex continuous constraint satisfaction problems \cite{FPSUZ17, franz2019jamming}, where the high dimensional limit is characterized by replica symmetry breaking \cite{MPV87}. In this case, one naturally expects that pure gradient descent dynamics goes to a stationary state where the system ages and drifts on a landscape of marginally stable minima \cite{cugliandolo1993analytical} \footnote{This has been found in numerical simulations in \cite{baity2018comparing}.}. The aging 
dynamics is controlled by an effective temperature that encodes for the roughness of the underlying landscape. However, it is well known that driving systems governed by glassy relaxation stops aging dynamics \cite{kurchan1997rheology, berthier2013non}.  Indeed, aging is essentially due to the progressive annealing in the landscape. More annealed systems surf on stationary points that are more and more stable and as a consequence their dynamics slows down. However, if the system is driven, the dynamics is renewed and aging stops. Based on the above considerations we may argue that both in the stationary state of the under-parametrized regime and in the early-time over-parametrized regime, the noise of SGD is a mixture of the noisy dynamics induced by the roughness of the underlining loss-landscape and the stochasticity induced by the algorithm itself. In this setting, it is useful to compare the dynamics to the one of complex driven systems such as low-temperature amorphous solids under deformation: the noise induces activated jumps between local minima, which can get further destabilized resulting in an avalanche dynamics \cite{nicolas2018deformation}. This may lead to power-law distributed jumps and connect with recent literature on L\'evy flights \cite{simsekli2019tail}. Further investigation on more complex models is needed to asses this phenomenology.
\acknowledgments 
We thank Lenka Zdeborov\'a for useful discussions. This work was
supported by ``Investissements d'Avenir'' LabExPALM (ANR-10-LABX-0039-PALM).

\appendix
\onecolumngrid
\section{Numerical procedure to extract the effective temperature}\label{appendix:teff}
In this section, we explain the procedure that we have used to estimate the effective temperature displayed in Fig.  1 and Fig.  2 of the main text. We treat the case of vanilla SGD and persistent SGD separately in the two following sections since they are characterized by some important differences.
\subsection{Vanilla stochastic gradient descent}
\begin{figure}[t!]
\centering
\includegraphics[scale=0.22]{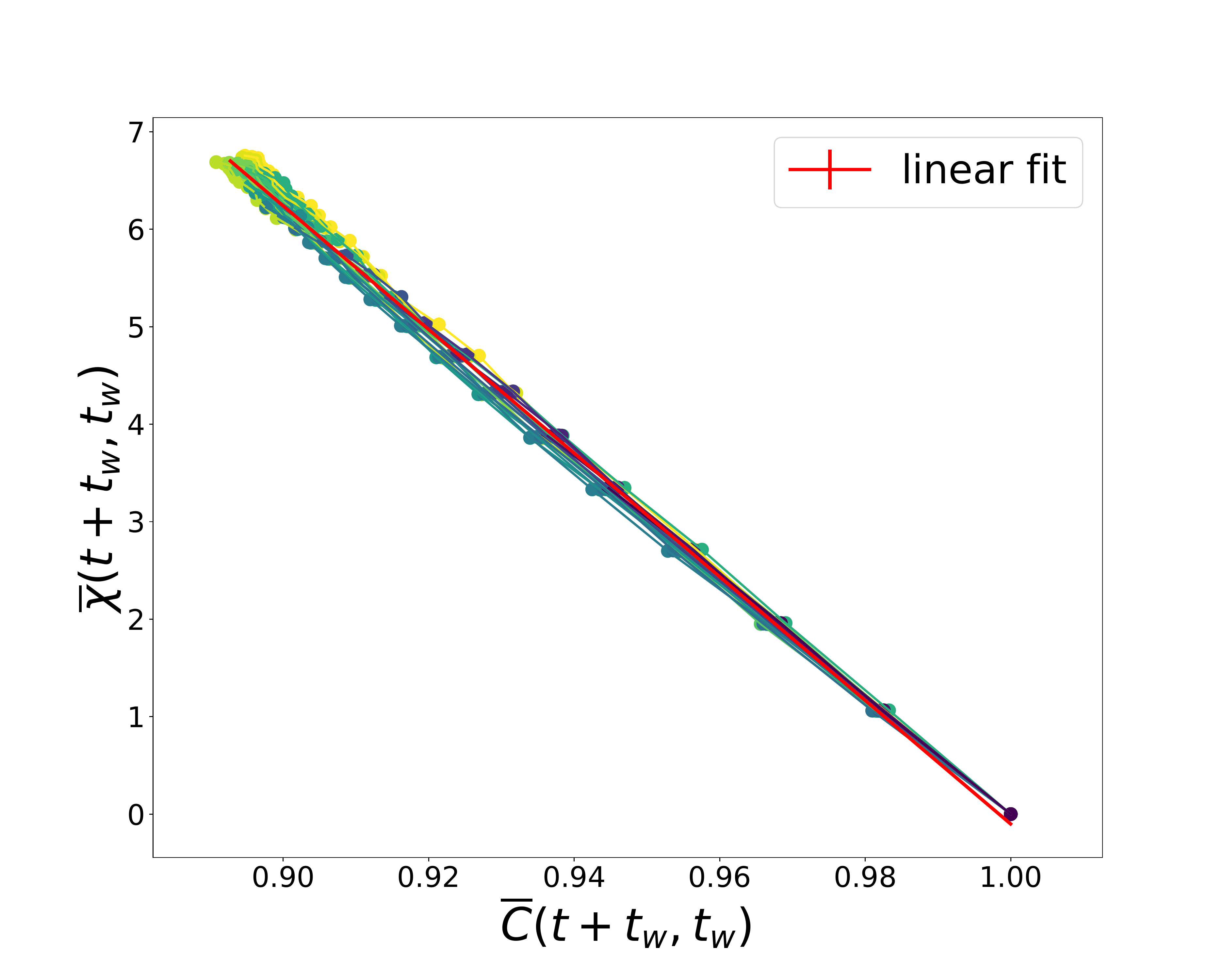}
\hspace{-2em}\includegraphics[scale=0.295]{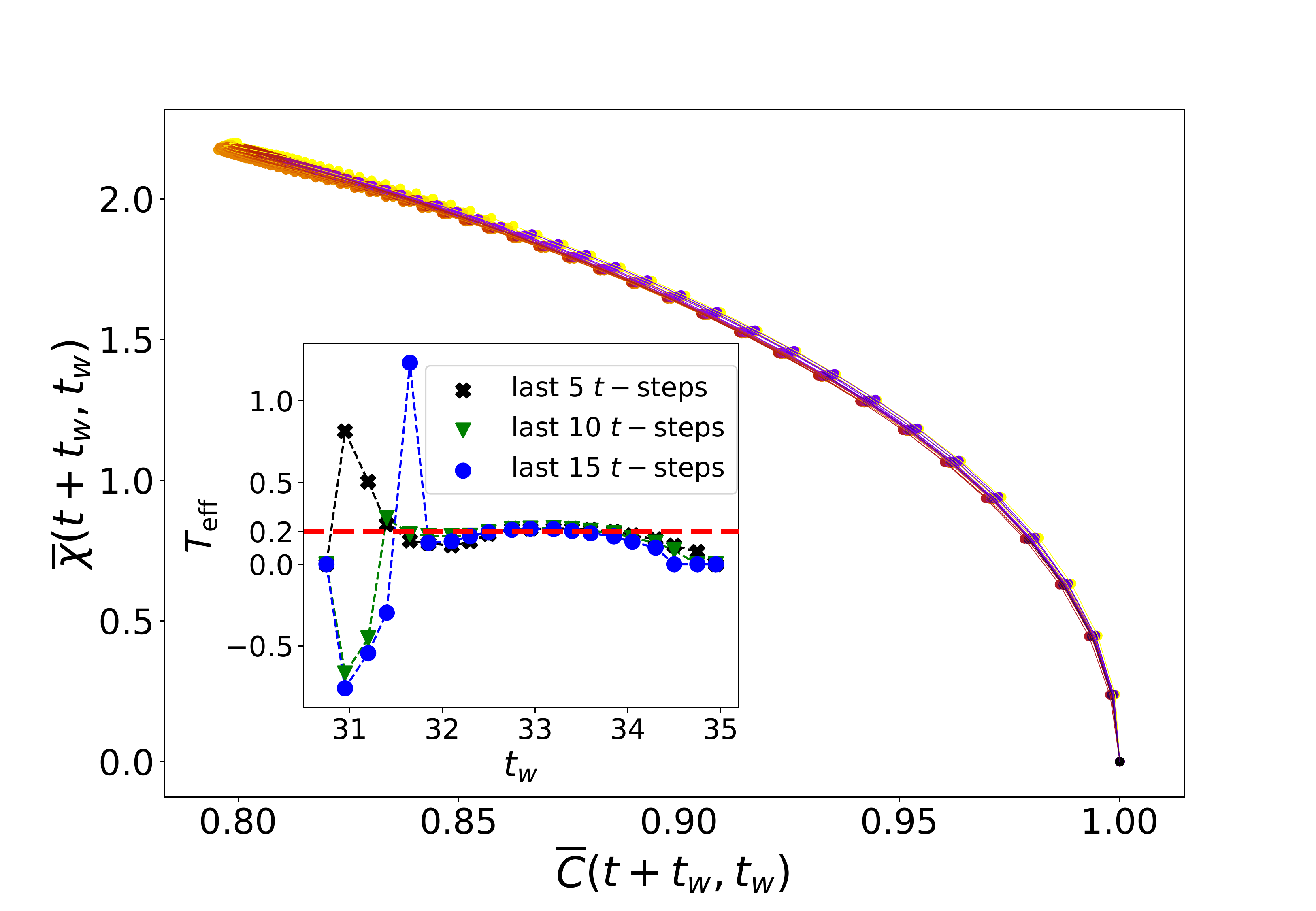}
\caption{\label{fig:compute_T}FDT plot for vanilla-SGD (left panel) and p-SGD (right panel). Different curves represent different values of waiting time $t_w$. Later waiting times $t_w$ are depicted with darker colors. Since the values of $t_w$ are such that the system is in the stationary state, all the curves are almost overlapping. At fixed $t_w$, each curve is a parametric plot with respect to $t\in[t_w,t_{\rm final}-t_w]$. The red line in the left panel represents the linear fit from which we compute the effective temperature $T_{\rm eff}$. For vanilla-SGD, we have fixed $\alpha=6,b=0.1,\dde t=0.1,\lambda=1,\Delta=1$ and we obtain $T_{\rm eff}=0.015\pm 0.00005$. The estimate of $T_{\rm eff}$ for p-SGD is shown in the inset of the right panel. For each value of $t_w$, we extrapolate the slope of the last $n$ points, corresponding to the last $n$ steps in time $t$, for different values of $n=5,10,15$. At large $t_w$, then $t=t_{\rm final}-t_w<b\tau$ and the slope tends to zero as $t_w$ goes to $t_{\rm final}$. We extrapolate the slope at smaller $t_w$, in the regime where $T_{\rm eff}$ is constant, as the value at which the curves for different $n$ converge. In the right panel, we consider $\alpha=8, b=0.2, \dde t=0.05, \Delta=1, \lambda=1$ and we obtain $T_{\rm eff}\approx 0.2$ for p-SGD.}
\end{figure}

Although vanilla-SGD is a discrete-time algorithm, we use the definition of Eq.~\eqref{eq:effFDT} to identify the effective temperature. The integration of the response function is performed numerically from the DMFT equations, therefore in practice a time-discretization is always needed. For vanilla-SGD, the variables $s_\mu(t)$ encoding the sampling process are i.i.d. at all times $t$. Therefore, in the stationary state, the FDT plot is a straight line and the effective temperature $T_{\rm eff}$ is a constant at all time differences. The most efficient way to extrapolate numerically the value of $T_{\rm eff}$ is hence to fit a line to all points $\{\left(\bar C(t+t_w,t+w),\bar\chi(t+t_w,t+w)\right)\}_{t\in [t_w,t_{\rm final}-t_w]}$ for a collection of large enough waiting times $t_w$ and time differences $t$ ranging between $t_w$ and the final time $t_{\rm final}$. This estimate is quite precise as it can be seen from the example displayed in the left panel of Fig.  \ref{fig:compute_T}.

\subsection{Persistent stochastic gradient descent}
\begin{figure}[t!]
\centering
\includegraphics[scale=0.35]{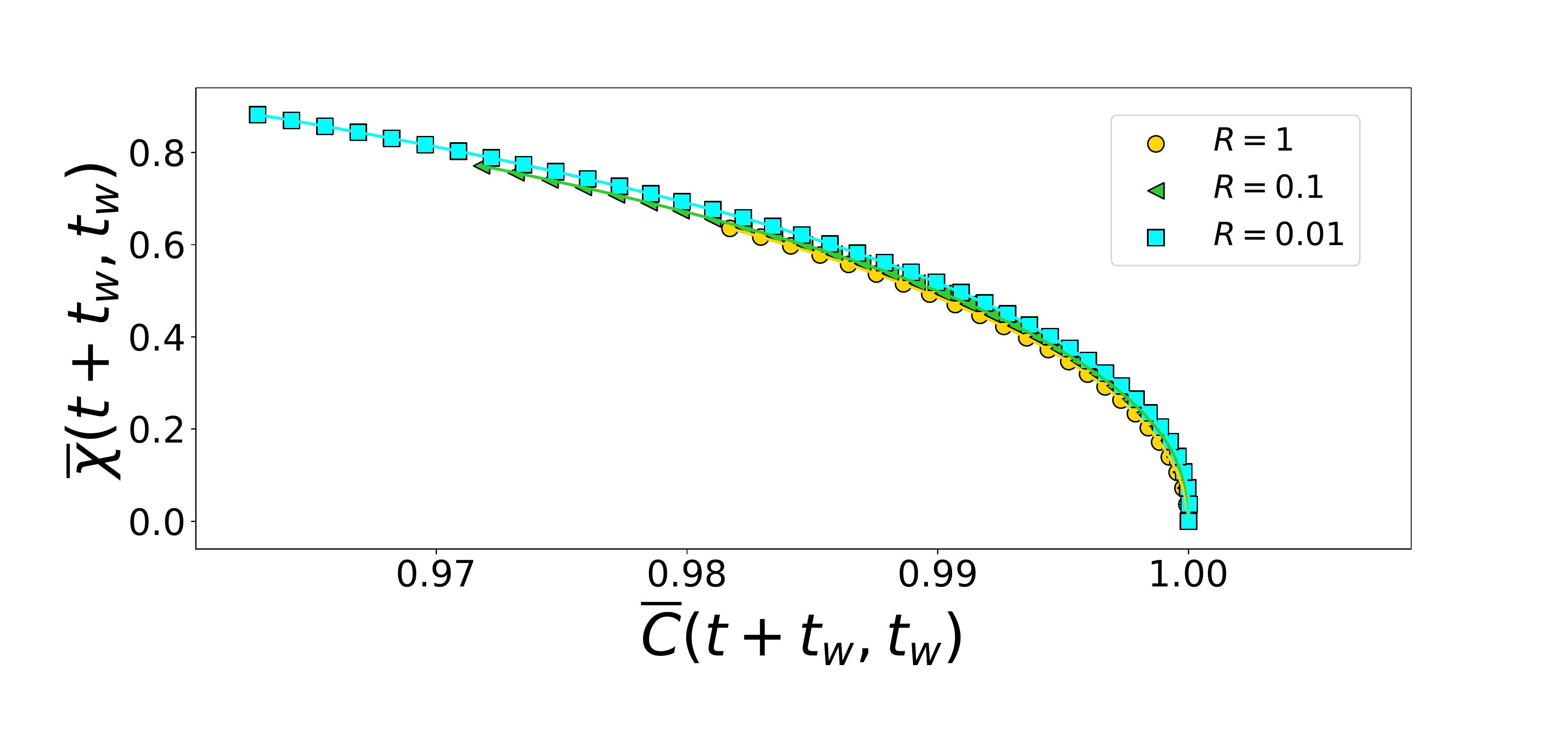}
\caption{\label{fig:init}FDT plot for p-SGD. We consider $\alpha=8$, $\Delta=1$, $\lambda=1$, where the classification problem is UNSAT and we use $\dd t=0.01$ and look at: different initialization variances $R=1,0.1,0.01$ and fixed $b=0.3$ and $\tau=2$. In all cases we obtain an estimate of the effective temperature $T_{\rm eff}\sim 0.05$. }
\end{figure}

In the case of p-SGD, the autocorrelation between a sampling variable at different times decays exponentially with the time difference $t>0$ at a rate that is given by the sum of the activation and deactivation rates $1/b\tau$, i.e., $\langle s(t_w)s(t+t_w)\rangle-b^2=b(1-b) \exp\left(-t/b\tau\right)$. This behavior is reflected by the fact that the effective temperature $T_{\rm eff}$ is not constant with the time difference. Instead, it is lower at small $t$ given the higher correlation between samples in the gradient, while it goes to a constant at $t$ larger than the typical decay time $b\tau$. This behavior can be clearly seen from Fig.  \ref{fig2} in the main text and is further elucidated in the right panel of Fig. \ref{fig:compute_T}. Therefore, in the case of  p-SGD one should actually define the effective temperature depending on the time difference $T_{\rm eff}(t)$ even in the stationary state. However, for simplicity we will refer to the constant $T_{\rm eff}(t>b\tau)$ as the effective temperature. This is motivated by the fact that $b\tau$ is usually small compared to the observation time and we are always able to observe this regime. The procedure through which we estimate $T_{\rm eff}$ for p-SGD is detailed in the right panel of Fig. \ref{fig:compute_T}. Finally, Fig. \ref{fig:init} shows that the effective temperature does not depend on the initialization variance $R$. Indeed, in the UNSAT phase the solution of the classification problem is unique and therefore the algorithm ends up rattling in the unique minimum with a noise strength given by $T_{\rm eff}$.

\section{Derivation of DMFT equations for two replicas of the system}\label{appendix:dmft}
In this section, we carry out the derivation of DMFT equations for two replicas of the system, corresponding to two different realizations of the gradient noise starting from the same initialization.
The derivation follows the lines of \cite{francesca2020dynamical}, with the difference that in this case we have two sets of dynamical weights variables, coupled by the disorder, i.e. the training dataset but whose dynamics is driven by two independent noises. Note that the dynamics of a single system can be derived from the single-replica quantities of the DMFT equations for two coupled replicas. Therefore, we only sketch the more general derivation for two coupled replicas.
The dynamical partition function has the form:
\begin{equation}
\begin{split}
Z_{\rm dyn}=\langle \int\left[\frac{\dde \ww^{(0)}}{(2\pi)^{N/2}}\, e^{-\frac 12 \Vert \ww^{(0)}\Vert^2}\right]\int_{\ww^1 (0)=\ww^2(0)=\ww^{(0)}} \mathcal{D}\ww^1(t)\mathcal{D}\ww^2(t)\\\times\prod_{j=1}^N\left(\delta\left(-\dot w^1_j(t)-\lambda w^1_j(t)-\sum_{\mu=1}^M s^1_\mu(t)y_\mu\ell'\left(y_\mu\frac{\ww^1(t)^\top \xx_\mu}{\sqrt{N}}\right) \frac{x_{\mu,j}}{\sqrt{N}}\right)\right.\\\left.\times\delta\left(-\dot w^2_j(t)-\lambda w^2_j(t)-\sum_{\mu=1}^M s^2_\mu(t)y_\mu\ell'\left(y_\mu\frac{\ww^2(t)^\top \xx_\mu}{\sqrt{N}}\right) \frac{x_{\mu,j}}{\sqrt{N}}\right)\right)\rangle\\
=\langle\int \mathcal{D}\ww^1(t)\mathcal{D}\ww^2(t)\mathcal{D}\hat\ww^1(t)\mathcal{D}\hat\ww^2(t) \,e^{S_{\rm dyn}}\rangle
\end{split},
 \label{Zdyn_replicas}
\end{equation}
where the brackets stand for the average over the realizations of the noise in the training set and in the SGD algorithms, while the average over the initial condition (the same for both replicas) is written explicitly. The dynamical action $S_{\rm dyn}$ is:
\begin{equation}
\begin{split}
S_{\rm dyn}=&\sum_{j=1}^N\int_0^{+\infty} \dde t \, i\hat w^1_j (t)\left(-\dot w_j^1(t) -\lambda w^1_j(t)- \sum_{\mu=1}^M s_\mu^1(t)y_\mu\ell'\left(y_\mu\frac{\ww^1(t)^\top \xx_\mu}{\sqrt{N}}\right) \frac{x_{\mu,j}}{\sqrt{N}}\right)\\&+\sum_{j=1}^N\int_0^{+\infty}\dde t  \, i\hat w^2_j (t)\left(-\dot w_j^2(t) -\lambda w^2_j(t)-\sum_{\mu=1}^M s_\mu^2(t)y_\mu\ell'\left(y_\mu\frac{\ww^2(t)^\top \xx_\mu}{\sqrt{N}}\right) \frac{x_{\mu,j}}{\sqrt{N}}\right).
\end{split}
\end{equation}
\paragraph*{\bf Supersymmetric formulation --} We introduce the super-fields\footnote[2]{We will always use a superscript $^{1,2}$ to indicate definitions that apply to either replica $1$ or replica $2$ separately.}: $\w^{1,2}(\sigma)=\w^{1,2}(t_\sigma)+i\theta_\sigma\bar\theta_\sigma\,\hat \w^{1,2}(t+\sigma)$, where $\theta_\sigma,\bar\theta_\sigma$ denote a couple of Grassman variables. We can then rewrite the dynamical action as:
\begin{equation}
\begin{split}
S_{\rm dyn}&=\int \dde \sigma \dde \gamma \,\mathcal{K}(\sigma,\gamma)\left(\ww^1(\sigma)^\top\ww^1(\gamma)+\ww^2(\sigma)^\top\ww^2(\gamma)\right)\\&-\sum_{\mu=1}^M\int \dde \sigma \left(s_\mu^1(\sigma)\ell (y_\mu h^1_\mu(\sigma))+s_\mu^2(\sigma)\ell (y_\mu, h^2_\mu(\sigma))\right),
\end{split}
\end{equation}
where we have defined $h^{1,2}_\mu(\sigma)=\ww^{1,2}(\sigma)^\top \xx_\mu/\sqrt{N}$,  $\mathcal{K}(\sigma,\gamma)=-\theta_\sigma\bar\theta_\sigma\partial_{t_\sigma}\delta(t_\gamma-t_\sigma)-\theta_\gamma\bar\theta_\gamma\partial_{t_\gamma}\delta(t_\sigma-t_\gamma)+\lambda\delta(\sigma,\gamma)$, and $\delta(\sigma,\gamma)=\delta(t_\gamma-t_\sigma)\left(\theta_\gamma\bar\theta_\gamma+\theta_\sigma\bar\theta_\sigma\right)$. After performing the average over the data noise, we find
\begin{equation}
Z_{\rm dyn}=\int\mathcal{D} \mathbf{Q}\,
\mathcal{D}\mathbf{m}\,e^{NS(\mathbf{Q},\mathbf{m})},
\end{equation}
where we have introduced the definitions
\begin{equation}
\begin{split}
m^{1,2}(\sigma)&=\frac{1}{N}\ww^{1,2}(\sigma)^\top \vv^*\,;\\
Q^{lk}(\sigma,\gamma)&=\frac{1}{N}\ww^l(\sigma)^\top \ww^k(\gamma),\qquad\; l,k\in\{1,2\}
\end{split}
\end{equation}
and
\begin{equation}
\begin{split}
S(\mathbf{Q},\mathbf{m})=&-\frac 12 \int \dde \sigma \dde \gamma\, \mathcal{K}(\sigma,\gamma)\left(Q^{11}(\sigma,\gamma)+Q^{22}(\sigma,\gamma)+m^1(\sigma)m^1(\gamma)+m^2(\sigma)m^2(\gamma)\right)\\&+\alpha\log\mathcal{Z}+\frac 12 \log\det\left(\mathbf{Q}(\sigma,\gamma)
\right),
\end{split}
\end{equation}
where we have performed a translation: $\mathbf{Q}\leftarrow\mathbf{Q}+\mathbf{m}\mathbf{m}^\top$, and for simplicity we have defined 
\begin{equation}
\label{eq:vec_notation}
\mathbf{Q}(\sigma,\gamma)=\left[\begin{array}{cccc}
Q^{11}(\sigma,\gamma) & \frac{1}{2}Q^{12}(\sigma,\gamma) \\
 \frac{1}{2}Q^{12}(\sigma,\gamma) & Q^{22}(\sigma,\gamma)
\end{array}\right],\quad \mathbf{m}(\sigma)=\left[\begin{array}{cc}
m^1(\sigma)\\ m^2(\sigma)
\end{array}\right],\quad \mathbf{h}(\sigma)=\left[\begin{array}{cc}
h^1(\sigma)\\ h^2(\sigma)
\end{array}\right],
\end{equation}
and
\begin{equation}
\begin{split}
\mathcal{Z}=\int \frac{\dde h_0}{\sqrt{2\pi}}e^{-h_0^2/2}\int \mathcal{D}h^1(\sigma)\mathcal{D}h^2(\sigma)\exp\left[-\frac{1}{2}\int\dde \sigma \dde \gamma \;\mathbf{h}(\sigma)^\top \mathbf{Q}^{-1}(\sigma,\gamma)\mathbf{h}(\gamma)\right.\\\left.-\int \dde \sigma  s^1(\sigma)\ell(y,\sqrt{\Delta}h^1(\sigma)+(y+\sqrt{\Delta}h_0)m^1(\sigma))\right.\\\left.-\int\dde\sigma s^2(\sigma)\ell(y,\sqrt{\Delta}h^2(\sigma)+(y+\sqrt{\Delta}h_0)m^2(\sigma))\right].
\end{split}
\end{equation}
after a translation $\mathbf{h}\leftarrow\mathbf{h}+y\mathbf{m}$ and a Hubbard-Stratonovich transformation.

\paragraph*{\bf Saddle point equations --} We can then evaluate the integral via a saddle point, leading to the following equations
\begin{equation}
\begin{split}
0&=-\mathcal{K}(\sigma,\gamma)+(Q^{-1})^{ll}(\sigma,\gamma)+2\alpha\frac{\delta}{\delta Q^{ll}(\sigma,\gamma)}\log\mathcal{Z},\qquad l\in\{1,2\}\\
0&=(Q^{-1})^{12}(\sigma,\gamma)+2\alpha\frac{\delta}{\delta Q^{12}(\sigma,\gamma)}\log\mathcal{Z},\\
0&=-\int\dde\gamma \mathcal{K}(\sigma,\gamma)m^l(\gamma)+\alpha\frac{\delta}{\delta m^{l}(\sigma)}\log\mathcal{Z}, \quad\qquad\qquad l\in\{1,2\}.
\end{split}
\end{equation}
The derivatives of $\mathcal{Z}$ are:
\begin{equation}
\begin{split}
2\alpha\frac{\delta}{\delta Q^{ll}(\sigma,\gamma)}\log\mathcal{Z}&=\alpha\Delta\left(\langle s^l(\sigma)s^l(\gamma)\ell'(yr^l(\sigma))\ell'(yr^l(\gamma))\rangle-\delta(\sigma,\gamma)\langle s^l(\sigma)\ell''(yr^l(\sigma))\rangle \right),\\
&=M^{ll}(\sigma,\gamma)-\delta(\sigma,\gamma)\delta\lambda^l(\sigma),\\
2\alpha\frac{\delta}{\delta Q^{12}(\sigma,\gamma)}\log\mathcal{Z}&=\alpha\Delta\langle s^1(\sigma)s^2(\gamma)\ell'(yr^1(\sigma))\ell'(yr^2(\gamma))\rangle=M^{12}(\sigma,\gamma),\\
\alpha\frac{\delta}{\delta m^{l}(\sigma)}\log\mathcal{Z}&=-\alpha\langle s^l(\sigma)(y+\sqrt{\Delta}h_0)y\ell'(yr^l(\sigma))\rangle=-\mu^l(\sigma),
\end{split}
\end{equation}
where we have used that $y^2=1$ and we have defined $r^l(\sigma)=\sqrt{\Delta}h^l(\sigma)+(y+\sqrt{\Delta}h_0)m^l(\sigma)$, for $l\in\{1,2\}$. At this point we can rewrite the averages in brackets as averages over the following stochastic processes:
\begin{equation}
\begin{split}
\partial_t h^1(t)=-(\lambda+\delta\lambda^1)h^1(t)-\sqrt{\Delta}s^1(t)y\ell'(yr^1(t))+\int_0^t\dde t'\, M^1_R(t,t')h^1(t')+\xi^1(t),\\
\partial_t h^2(t)=-(\lambda+\delta\lambda^2)h^2(t)-\sqrt{\Delta}s^2(t)y\ell'(yr^2(t))+\int_0^t\dde t'\, M^2_R(t,t')h^2(t')+\xi^2(t),\label{eq:h_process}
\end{split}
\end{equation}
coupled by the variable $h_0$ and the correlation of the effective Gaussian noise, given by
\begin{equation}
\begin{split}
\langle\xi^l(t)\xi^k(t')\rangle&=M_C^{lk}(t,t'),\\M_C^{lk}(t,t')&=\alpha\Delta\langle s^l(t)s^k(t')\ell'(yr^l(t))\ell'(yr^k(t'))\rangle\qquad l,k\in\{1,2\},
\end{split}
\end{equation} 
while the memory kernel is:
\begin{equation}
M^l_R(t,t')=\alpha\Delta\frac{\delta}{\delta P^l(t')}\langle s^l(t)y\ell'(yr^l(t))\rangle\biggr\rvert_{P^l(t')=0}.
\end{equation}
The magnetizations evolve accordinge to the ODEs:
\begin{equation}
\partial_t m^l(t)=-\lambda m^l(t)-\mu^l(t),\qquad l\in\{1,2\}.\label{eq:ODE_m}
\end{equation}
We have also introduced the auxiliary functions:
\begin{equation}
\begin{split}
\delta\lambda^l(t)&=\alpha\Delta\langle s(t)\ell''(yr^l(t))  \rangle,\\
\mu^l(t)&=\alpha\langle s(t) \left(y+\sqrt{\Delta}h_0\right)\ell'(yr^l(t))\rangle.
\end{split}
\end{equation}
\paragraph*{\bf Correlation and response functions --} At this point, we can recover the correlation and response functions from the time-dependent overlap $Q^{lk}$, where we have to take into account the translation performed to compute the integrals. We obtain
\begin{equation}
\begin{split}
Q^{ll}(\sigma,\nu)&=C^{ll}(t_\sigma,t_\nu)-m^l(t_\sigma)m^l(t_\nu)+\theta_\sigma\bar\theta_\sigma R^l(t_\nu,t_\sigma)+\theta_\nu\bar\theta_\nu R^l(t_\sigma,t_\nu),\qquad l\in\{1,2\}\\
Q^{12}(\sigma,\nu)&=C^{12}(t_\sigma,t_\nu)-m^1(t_\sigma)m^2(t_\nu),\\
M^{lk}(\nu,\gamma)&=M^{lk}_C(t_\nu,t_\gamma)+\delta_{lk}\left(\theta_\nu\bar\theta_\nu M^l_R(t_\gamma,t_\nu)+\theta_\gamma\bar\theta_\gamma M^k_R(t_\nu,t_\gamma)\right),\qquad\qquad l,k\in\{1,2\}.
\end{split}
\end{equation}
We can then compute the equations for correlation and response functions from the following closure relations
\begin{equation}
\begin{split}
\delta(t_\sigma-t_\gamma)\left(\theta_\sigma\bar\theta_\sigma+\theta_\gamma\bar\theta_\gamma\right)=\int\dde\nu \left[Q^{ll}(\sigma,\nu)(Q^{-1})^{ll}(\nu,\gamma)+Q^{12}(\sigma,\nu)(Q^{-1})^{12}(\nu,\gamma)\right]\\=
\int\dde\nu\left[Q^{ll}(\sigma,\nu)\left(\mathcal{K}(\nu,\gamma)+\delta(\nu,\gamma)\delta\lambda^l(\nu)-M^{ll}(\nu,\gamma)\right)-Q^{12}(\sigma,\nu)M^{12}(\nu,\gamma)\right],\\
0=
\int\dde\nu \left[Q^{ll}(\sigma,\nu)(Q^{-1})^{12}(\nu,\gamma)+Q^{12}(\sigma,\nu)(Q^{-1})^{kk}(\nu,\gamma)\right]\\=
\int\dde\nu\left[-Q^{ll}(\sigma,\nu)M^{12}(\nu,\gamma)+Q^{12}(\sigma,\nu)\left(\mathcal{K}(\nu,\gamma)+\delta(\nu,\gamma)\delta\lambda^k(\nu)-M^{kk}(\nu,\gamma)\right)\right],
\end{split}
\end{equation}
where $l\neq k$, $l,k\in\{1,2\}$. We can then derive the equations for the correlation and response functions:
\begin{equation}
\begin{split}
\partial_{t'}C^{ll}(t,t')=&-\tilde \lambda^l(t')C^{ll}(t,t')+\int_0^t\dde s \, R^l(t,s)M_C^{ll}(s,t')+\int_0^{t'}\dde s \, C^{ll}(t,s)M_R^l(t',s)\\
&-m^l(t)\left(\mu^l(t')-\delta\lambda^l(t')m^l(t')+\int_0^{t'}\dde s\, M^l_R(t',s)m^l(s)\right),\qquad\forall t\neq t'\,,\,l\in\{1,2\}\\
\partial_{t'}C^{12}(t,t')=&-\tilde\lambda^l(t') C^{12}(t,t')+\int_0^t \dde s\,R^l(t,s)M_C^{12}(s,t')+\int_0^{t'}\dde s\, C^{12}(t,s)M^k_R(t',s)\\&-m^l(t)\left(\mu^k(t')-\delta\lambda(t')m^k(t')+\int_0^{t'}\dde s \, M^k_R(t',s)m^k(s)\right),\qquad\forall t\neq t'\,,\,l\neq k\\
\partial_{t'}R^l(t',t)=&-\tilde \lambda^l(t')R^l(t',t)+\delta(t'-t)+\int_t^{t'}\dde s \, M^l_R(t',s)R^l(s,t).
\end{split}\label{correlations}
\end{equation}
Note that, for the symmetry of the definition $Q^{12}=Q^{21},\,C^{12}=C^{21}$. Moreover, since the two replicas are statistically indistinguishable, also all singe-replica quantities should match.  Note that all single-replicas quantities can be used to compute the correlation and response and hence the effective temperature.\\
\paragraph*{\bf Numerical integration of the equations --} The complexity of Eqs. \eqref{eq:h_process} prevents us from solving them analytically. Instead, we resort to a numerical iterative method to integrate them. The procedure is analogous to the one applied in \cite{francesca2020dynamical}: we start from random guesses for the kernels and auxiliary functions; we plug these guesses in Eqs. \eqref{eq:h_process} to generate multiple realizations of the stochastic processes; we use this realizations to update the kernels/auxiliary functions ($M_C, M_R, \hat\lambda,\mu$) with new estimates; we update the value of $m$ integrating Eq. \eqref{eq:ODE_m}; we repeat until convergence. Once convergence has been reached, we use the final values of kernels, auxiliary functions and to compute correlation and responses from Eq.s \eqref{correlations}. 
\section{Computation of the fraction of support vectors at late times}\label{appendix:sv}
\begin{figure}[t!]
\centering
\includegraphics[scale=0.34]{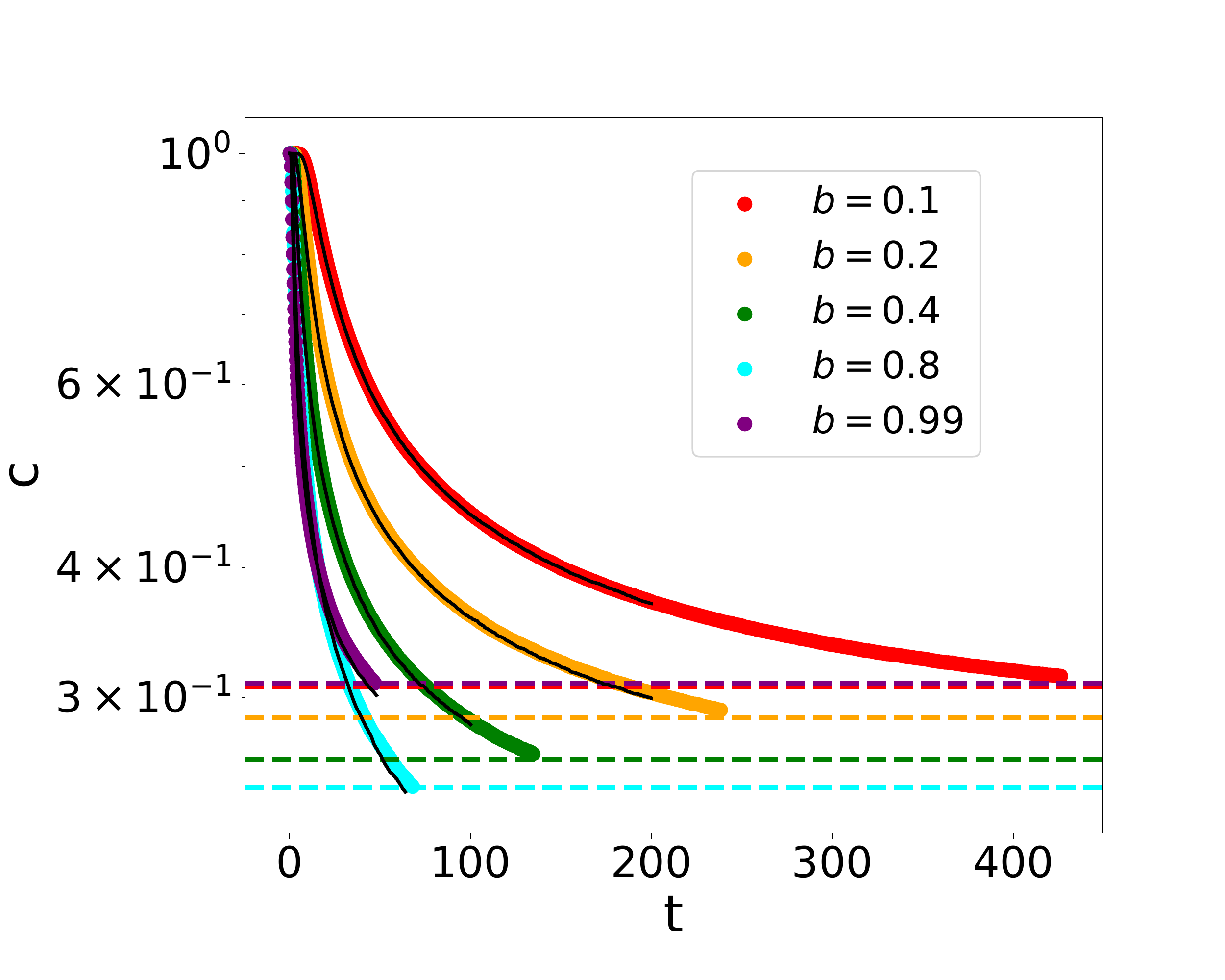}
\includegraphics[scale=0.34]{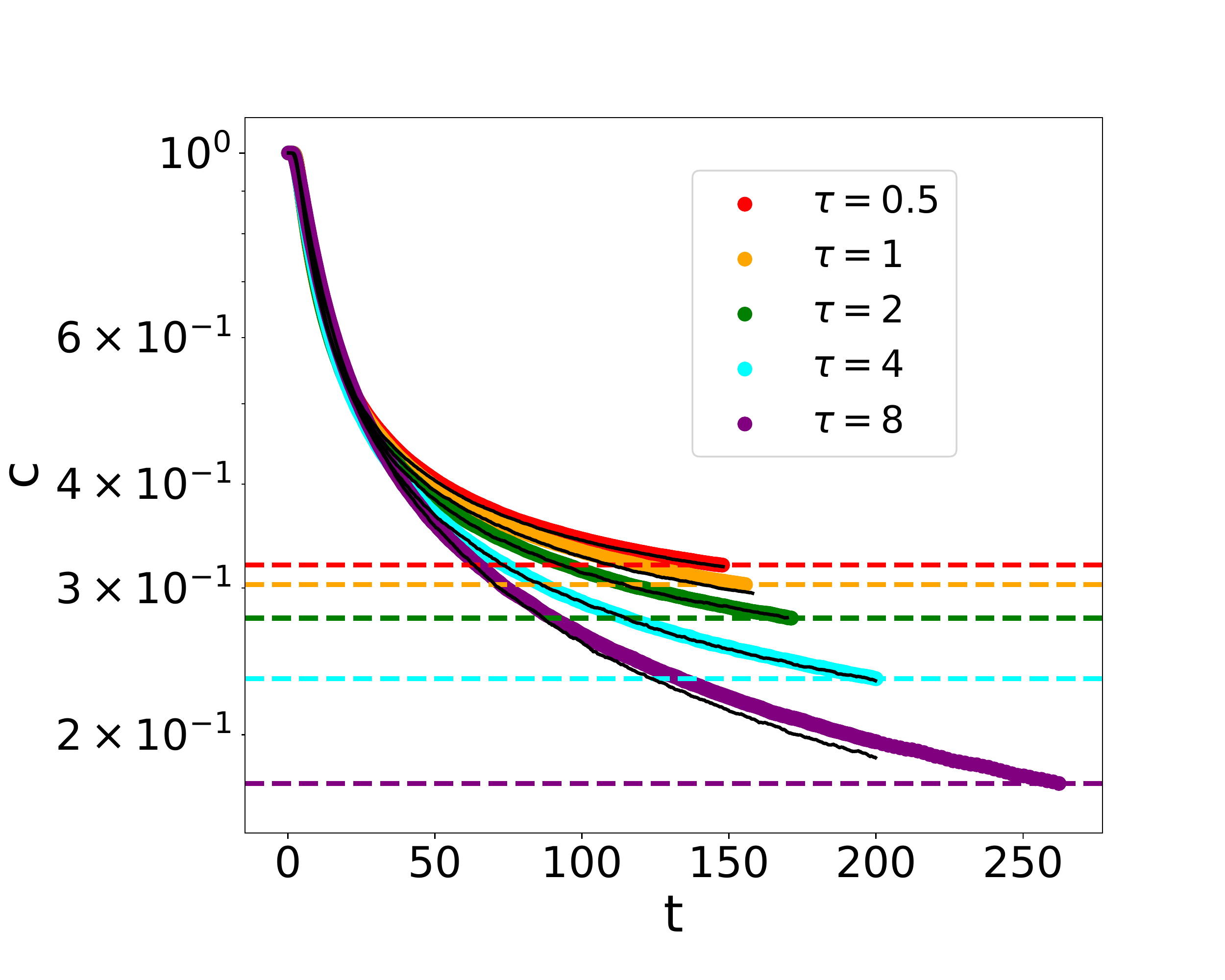}
\caption{\label{sv_estimate} Fraction of support vectors as a function of time for different values of batch size $b$ and fixed persistence time $\tau=2$ (left panel), and for different values of $\tau$ at fixed $b$ (right panel). The horizontal dashed line mark the stopping times at which we have computed the value of $c_0$. The average energy has reached $10^{-9}-10^{-10}$ for all parameter settings. The full lines correspond to the DMFT solution.}
\end{figure}

In this section, we explain the procedure that we have used to extract the typical fraction of support vectors of a given solution, We consider the following observable:
\begin{equation}
c(\w)=\frac{1}{M}\sum_{\mu=1}^M\Theta\left(\kappa-\frac{y_\mu}{\sqrt{N}}\ww^\top\xx_\mu\right).
\end{equation}
At the beginning of training, the value of $c$ will depend on the relative values of the initialization variance $R$ and the margin $\kappa$. However, during training, this quantity is decreased as more and more samples are correctly classified. As explained in the main text, due to the time discretization in the algorithms, the dynamics will never stop exactly at the border of the solutions set that it would reach in continuous time. Therefore, we need to fix a criterion to stop the simulation in order to characterize the border close to a given solution. Along the lines of \cite{hwang2020force}, where the authors consider the evolution of $c$ in full-batch gradient descent, we fix a threshold on the stochastic gradient, that we rescale by the batch size in order to compare the different algorithms on equal footing. Our stopping criterion is therefore:
\begin{equation}
\parallel \tilde \nabla _{\bf w}^{{\bf s}(t)}\mathcal{L}({\bf w}(t))\parallel_2^2/bN \leq 10^{-10}.
\end{equation}
We then compute the value of $c$ at the stopping time, averaged over multiple runs of the simulations at a given pair $(b,\tau)$ of persistence time and batch size. The fraction of support vectors is decreasing in time and approaches a finite value at late times. We derive the late-time limit of the average distance between two replicas of the system $d_0=\lim_{t\rightarrow\infty}d(t)$ in a similar manner.  Our results are displayed in Fig. \ref{sv_estimate}. We find that the DMFT is in good agreement with the numerical simulations.

\section{Numerics on the zero-learning-rate limit}
\label{appendix:dt_zero}
\begin{figure*}[t!]
\centering
\begin{subfigure}[t]{.49\textwidth}
\centering
\includegraphics[width=\textwidth]{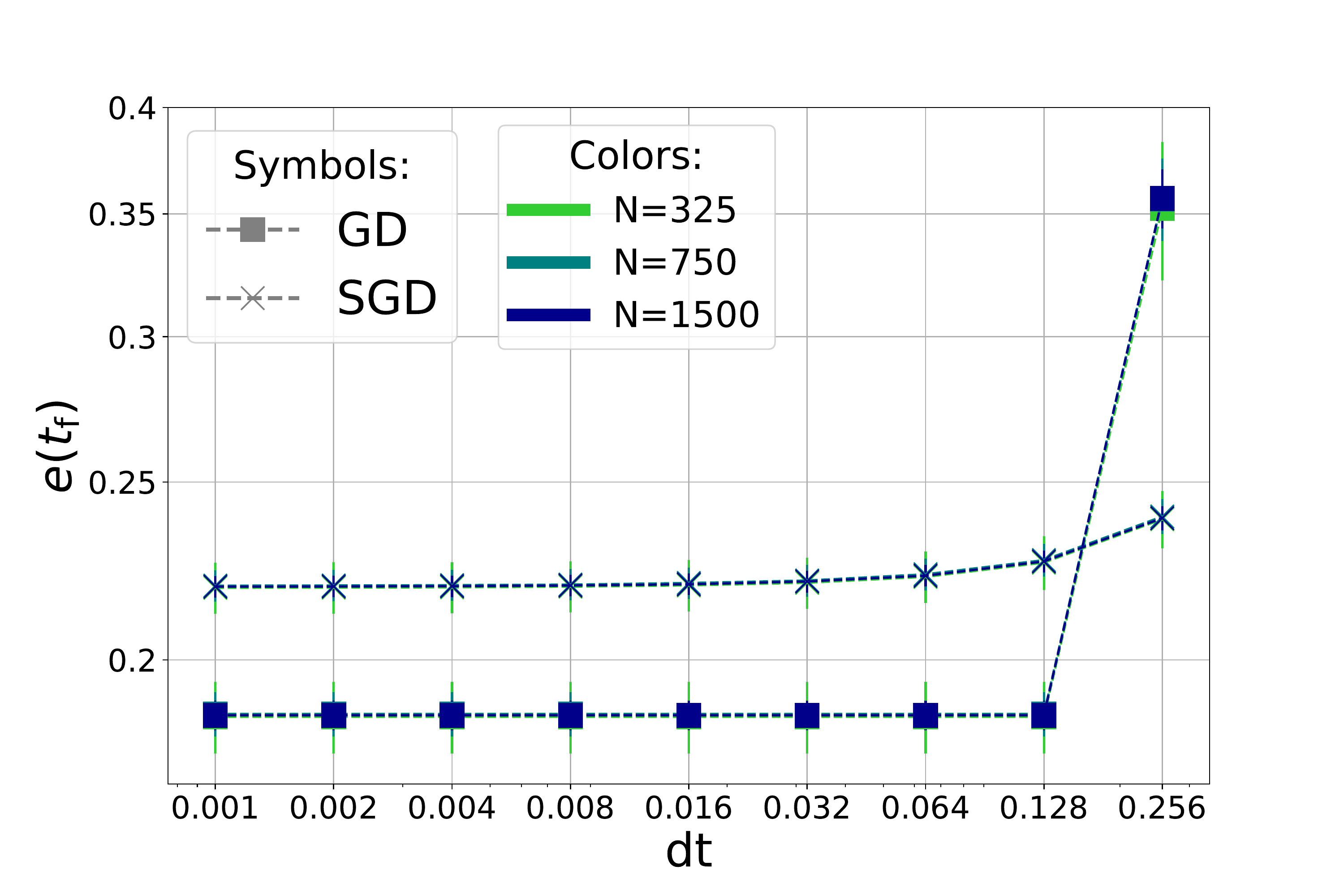}
\caption{Average \textbf{training loss} at the end of training.}
\end{subfigure}
\begin{subfigure}[t]{.49\textwidth}
\centering
\includegraphics[width=\textwidth]{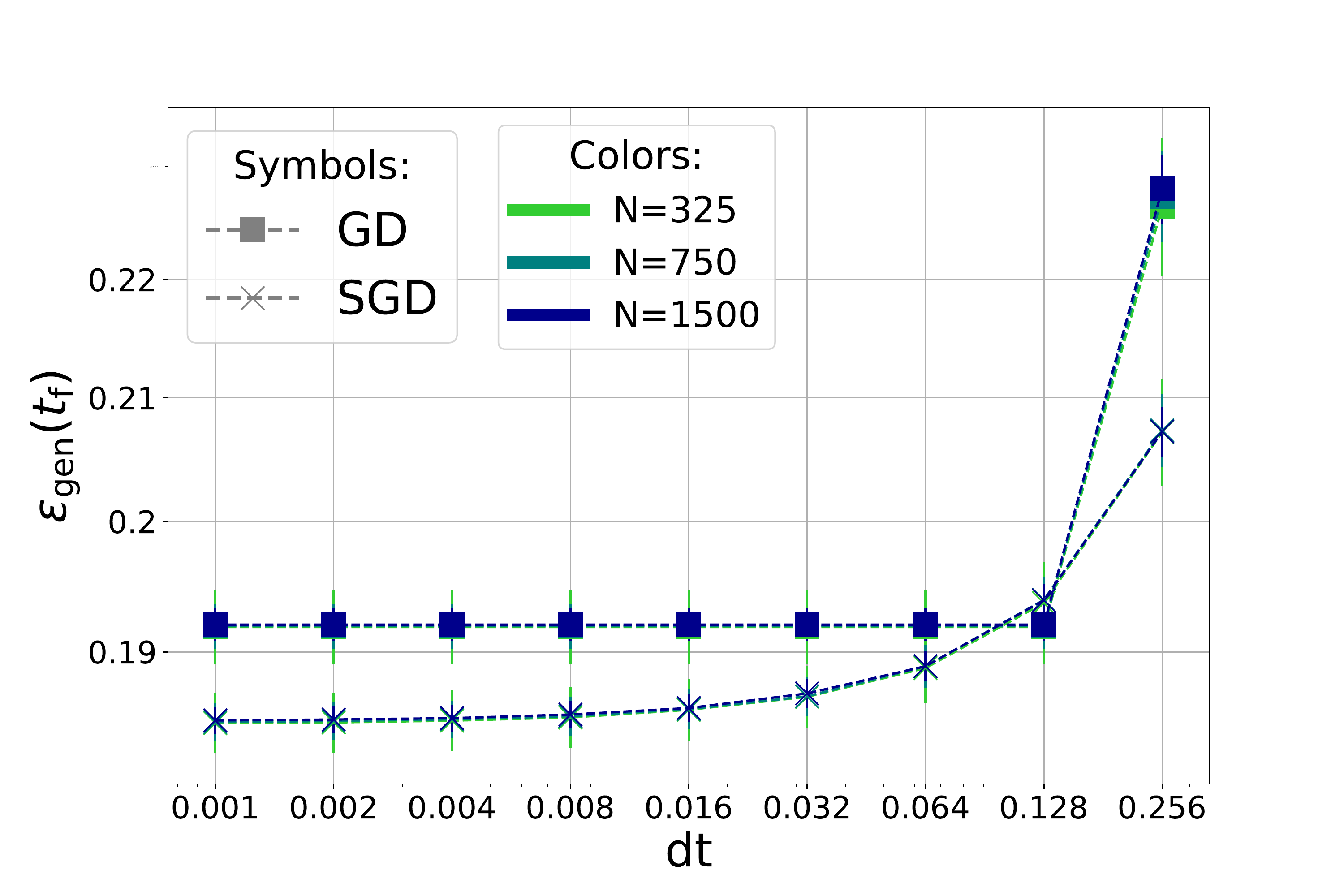}
\caption{Average \textbf{generalization error} at the end of training.}
\end{subfigure}
\begin{subfigure}[t]{.49\textwidth}
\centering
\includegraphics[width=\textwidth]{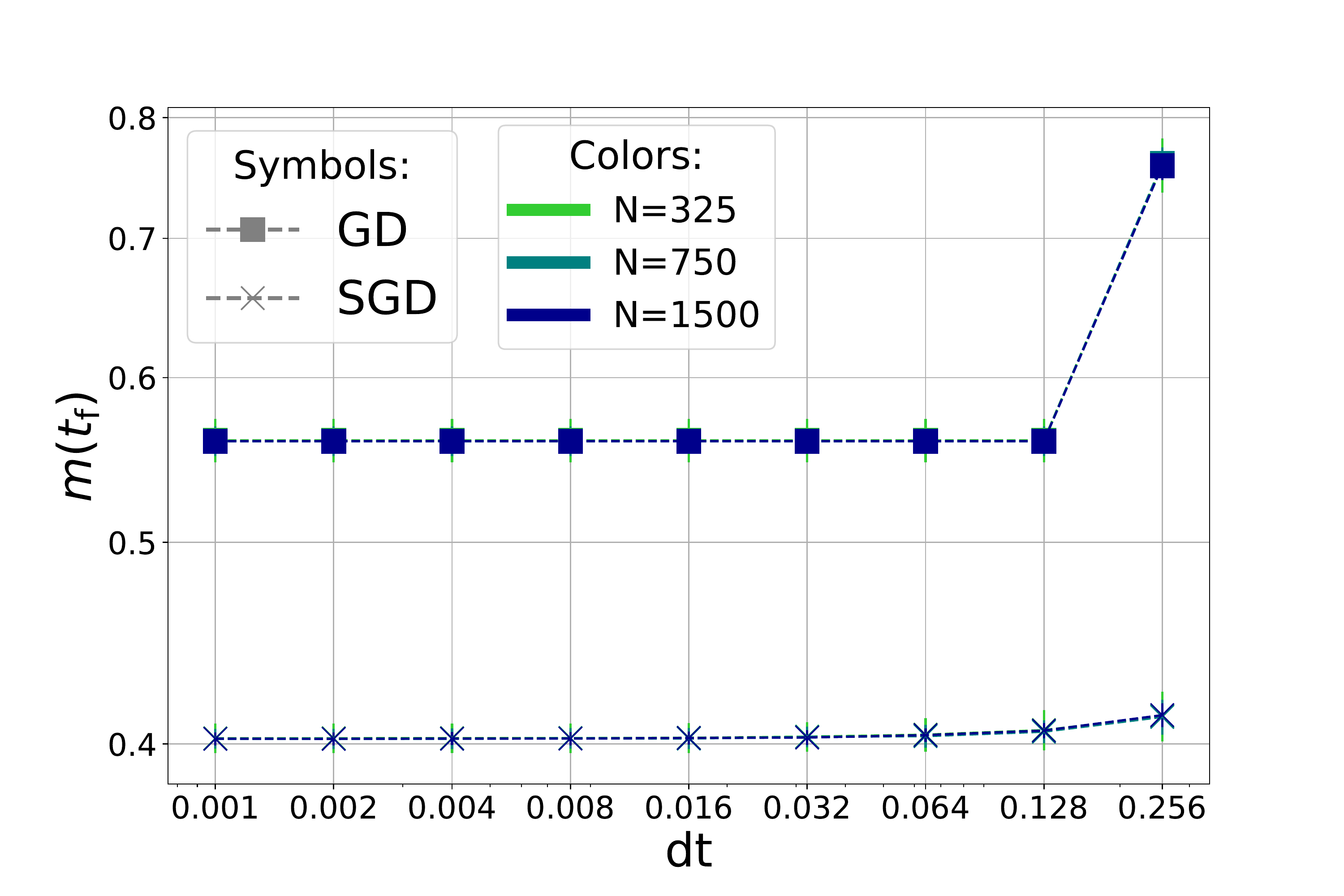}
\caption{Average \textbf{magnetization} at the end of training.}
\end{subfigure}
\begin{subfigure}[t]{.49\textwidth}
\centering
\includegraphics[width=\textwidth]{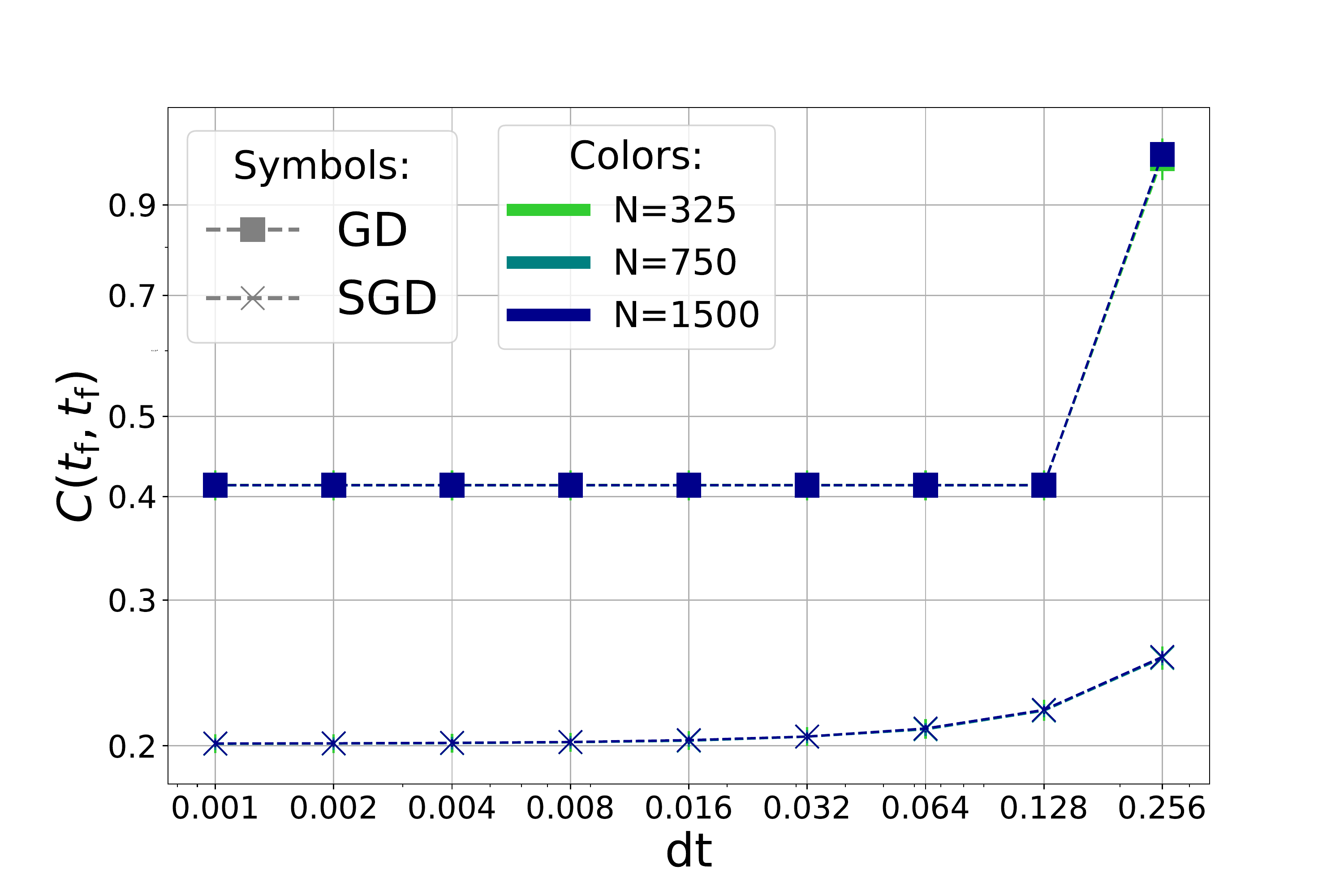}
\caption{Average \textbf{squared norm} at the end of training.}
\end{subfigure}
\caption{Numerical simulations on the behavior of GD and SGD at the end of training, as a function of the learning rate $\dde t$. We fix the parameter space such that the problem lies in the UNSAT phase ($\alpha=6$, $\lambda=1$, $\Delta=1$). The squares mark the behavior of GD while the crosses represent SGD at batch size $b=0.3$. Different colors mark different system sizes: $N=325$ (green), $N=750$ (teal),  $N=1500$ (blue). The simulations are averaged over $150$ realizations of the input data, the initialization and the sampling noise.\label{fig:small_dt}}
\end{figure*}

In this section we show the results of numerical simulations investigating the behavior of the GD and SGD algorithms in the limit of vanishing learning rate $\dde t\rightarrow 0^+$.  Fig.~\ref{fig:small_dt} displays the final values reached by the training loss (top-left panel), the generalization error (top-right panel), the magnetization (bottom-left panel) and the squared norm (bottom-right panel) for SGD (cross symbols) and GD (square symbols) at the end of training. The results are averaged over $150$ realizations of the initial condition, the input data and the sampling noise. We observe that the results are stable with respect to different system sizes $N=325,750,1500$, marked by different colors. These simulations strongly suggest that, in the problem under consideration, the high-dimensional limit of the dynamics of SGD remains different from that of GD, even in the limit of vanishing learning rate. Indeed, all the observables in Fig.~\ref{fig:small_dt}, i.e., those that determine the algorithmic performance, reach different plateaus for SGD and GD as $\dde t \rightarrow 0^+$. This behavior is at variance with that observed in other high-dimensional problems, e.g., phase retrieval \cite{Mignacco_2021}.

\section{Average distance between two replicas in the UNSAT phase}
\label{appendix:distance_unsat}
\begin{figure*}
\begin{subfigure}[t]{.49\textwidth}
\includegraphics[width=\textwidth]{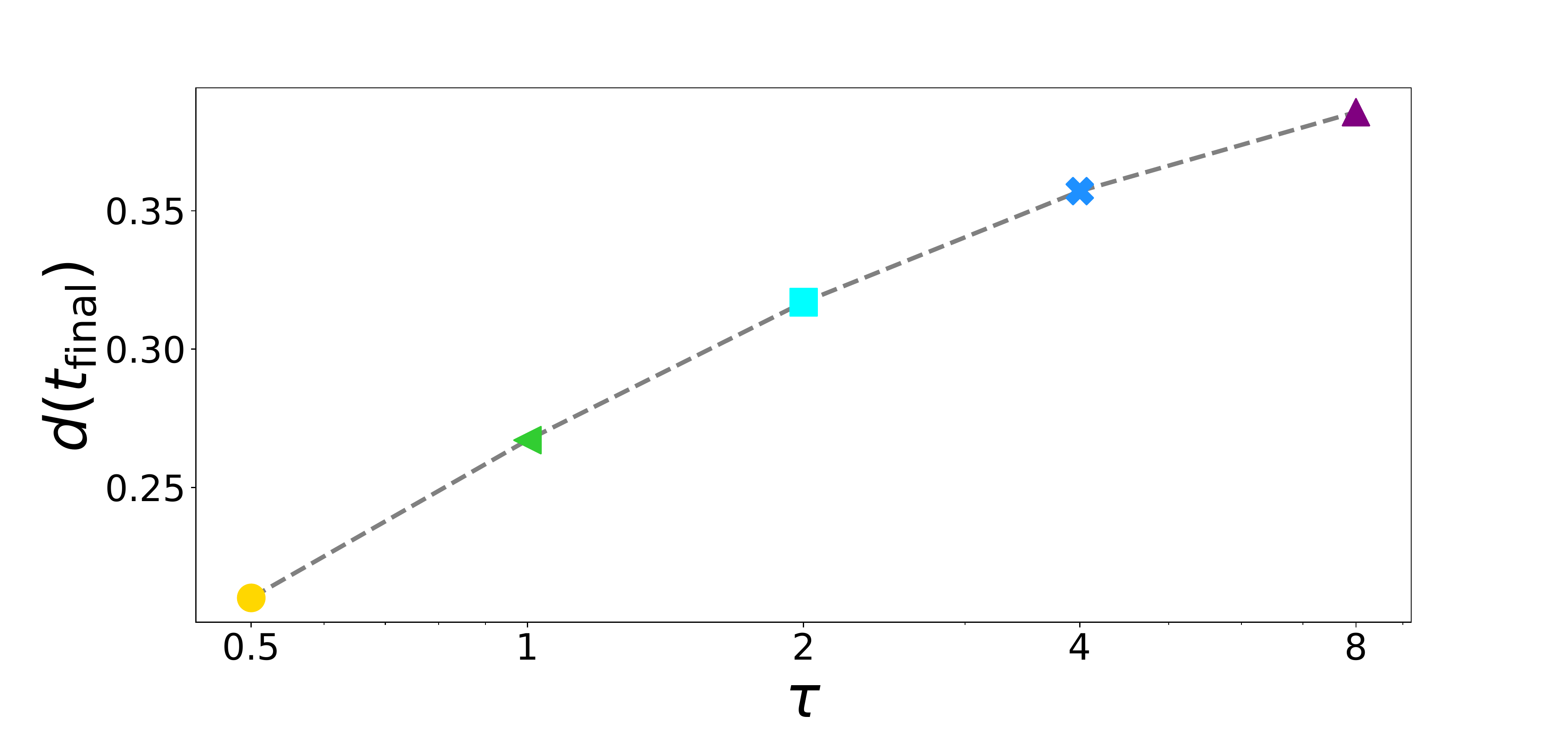}
\caption{Changing persistence time $\tau=0.5,1,2,4,8$ at fixed batch size $b=0.3$.}
\end{subfigure}
\begin{subfigure}[t]{.49\textwidth}
\includegraphics[width=\textwidth]{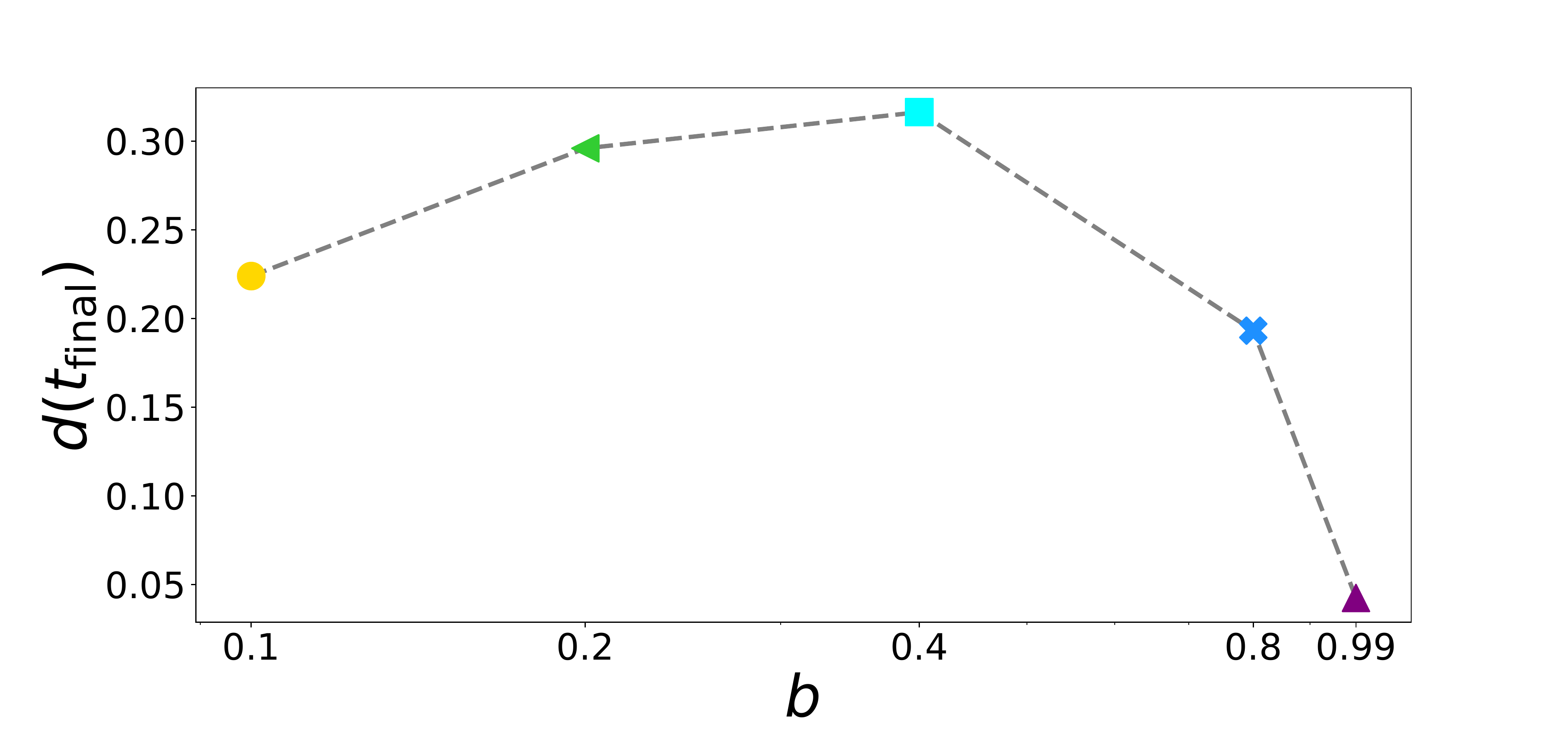}
\caption{Changing batch size $b=0.1,0.2,0.4,0.8,0.99$ at fixed persistence time $\tau=2$.}
\end{subfigure}
\caption{\label{fig:distance_unsat} Average final distance $d(t_{\rm final})$ between two copies of the p-SGD dynamics. The results are obtained by numerical simulations at size $N=750$, averaged over $100$ seeds. The other parameters are the same as in Fig. \ref{fig2}, such that the problem lies in the UNSAT phase: $\alpha=8,\Delta=1,\lambda=1$. The learning rate is ${\rm d}t=0.05$. }
\end{figure*}

In Fig. \ref{fig:distance_unsat}, we plot the average final distance between two replicas of the p-SGD dynamics in the UNSAT phase.The replicas share the same initialization, landscape and noise hyper-parameters ($\tau,b,\eta$) but are subjected to two different sampling histories. We observe that the qualitative behavior of this alternative measure of the algorithmic noise is consistent with the one of the effective temperature in the same region of the parameter space.

\clearpage
\bibliography{refs}
\end{document}